\documentclass[aps,twocolumn,superscriptaddress,shortbibliography]{revtex4-1}

\usepackage{pdfpages}
\usepackage{amsmath,amssymb}
\usepackage{graphicx}
\usepackage{hyperref}

\newcommand{\ex}[1]{\langle{#1}\rangle}
\newcommand{\tr}{\mathrm{Tr}}
\newcommand{\sss}{\mathcal{S}}
\newcommand{\heff}{H_{\mathrm{eff}}}

\makeatletter
\AtBeginDocument{\let\LS@rot\@undefined}
\makeatother

\begin{document}
\title{Self-learning Monte Carlo with Deep Neural Networks}

\author{Huitao Shen}
\email{huitao@mit.edu}
\affiliation{Department of Physics, Massachusetts Institute of Technology, Cambridge, Massachusetts 02139, USA}
\author{Junwei Liu}
\email{liuj@ust.hk}
\affiliation{Department of Physics, Massachusetts Institute of Technology, Cambridge, Massachusetts 02139, USA}
\affiliation{Department of Physics, Hong Kong University of Science and Technology, Clear Water Bay, Hong Kong, China}
\author{Liang Fu}
\affiliation{Department of Physics, Massachusetts Institute of Technology, Cambridge, Massachusetts 02139, USA}

\begin{abstract}
Self-learning Monte Carlo (SLMC) method is a general algorithm to speedup MC simulations. Its efficiency has been demonstrated in various systems by introducing an effective model to propose global moves in the configuration space. In this paper, we show that deep neural networks can be naturally incorporated into SLMC, and without any prior knowledge, can learn the original model accurately and efficiently. Demonstrated in quantum impurity models, we reduce the complexity for a local update from $ \mathcal{O}(\beta^2) $ in Hirsch-Fye algorithm to $ \mathcal{O}(\beta \ln \beta) $, which is a significant speedup especially for systems at low temperatures.
\end{abstract}

\maketitle

As an unbiased method, Monte Carlo (MC) simulation plays an important role in understanding condensed matter systems. Although great successes have been made in the past several decades, there are still many interesting systems that are practically beyond the capability of conventional MC methods, due to the strong autocorrelation of local updates or due to the heavy computational cost of a single local update. In the midst of recent developments of machine learning techniques in physics \cite{Carrasquilla2017,Wang2016a,Tanaka2017a,Ohtsuki2017,VanNieuwenburg2017,Zhang2017a,Mills2017a,Wetzel2017,Hu2017a,Schindler2017,Lu2017,Cristoforetti2017,Wang2017,Zhang2017,Rao2017,Yoshioka2017,Huembeli2017}, a general method called ``Self-learning Monte Carlo (SLMC)'' was introduced to reduce or solve these problems, first in classical statistical mechanics models \cite{Liu2017, Huang2017}, later extended to classical spin-fermion models \cite{Liu2017a}, determinantal quantum Monte Carlo (DQMC) \cite{Xu2017}, continuous-time quantum Monte Carlo \cite{Huang2017a,Nagai2017} and hybrid Monte Carlo \cite{Tanaka2017}. Recently, it helped understand itinerant quantum critical point by setting up a new record of system size in DQMC simulations \cite{Liu2017b}.

Designed under the philosophy of ``first learn, then earn'', the central ingredient of SLMC is an effective model that is trained to resemble the dynamics of the original model. The advantage of SLMC is two-fold. First, simulating the effective model is much faster, which enables the machine to propose global moves to accelerate MC simulations on the original model. Second, the effective model can directly reveal the underlying physics, such as the RKKY interaction in the double-exchange model \cite{Liu2017a} and the localized spin-spin imaginary-time correlation \cite{Nagai2017}. We note that there have been many previous works incorporating effective potentials or proposing various kinds of global moves to improve Monte Carlo simulation efficiency \cite{PhysRevLett.82.4745,doi:10.1002/cphc.200400587,PhysRevE.66.046701,PhysRevE.78.056707,PhysRevE.82.046710}.

The efficiency of SLMC depends on the accuracy of the effective model, which is usually invented based on the human understanding of the original system \cite{Liu2017,Liu2017a,Xu2017,Huang2017a,Nagai2017,Liu2017b}. To further extend SLMC to complex systems where an accurate effective model is difficult to write down, in this work, we employ deep neural networks (DNN) as effective models in SLMC. Instead of treating neural networks as black boxes with a huge number of parameters and training them blindly, we show how to design highly efficient neural networks that respect the symmetry of the system, with very few parameters yet capturing the dynamics of the original model quantitatively. The generality of this approach is guaranteed by the mathematical fact that DNNs are able to accurately approximate any continuous functions given enough fitting parameters \cite{Cybenko1989,Hornik1991}. Practically, our DNNs can be trained with ease using back-propagation-based algorithms \cite{Goodfellow2016}, and can be directly evaluated in dedicated hardwares \cite{Jouppi2017}. Compared with other machine learning models in SLMC such as the restricted Boltzmann machine \cite{Huang2017,Tanaka2017}, which can be regarded as a fully-connected neural network with one hidden layer, our DNNs have greater expressibility (more hidden layers) and flexibility (respecting the symmetry of the model).

As a concrete example, we demonstrate SLMC with DNNs on quantum impurity models.
In the following, we first review SLMC for fermion systems. We then implement the simplest neural networks and test their performances. Next, we show how the visualization of these networks helps design a more sophisticated convolutional neural network that is more accurate and efficient. Finally, we discuss the complexity of our algorithm. 

\textit{SLMC for Fermions}
For an interacting fermion system, the partition function is given by $ Z=\tr{[e^{-\beta \hat{H}_f}]} $, where $ \beta=1/T $ is the inverse temperature, and the trace is over the grand-canonical ensemble. One often applies the Trotter decomposition $ e^{-\beta \hat{H}_f}=\prod_{i=1}^L e^{-\Delta\tau \hat{H}_f}  $, $ \Delta\tau=\beta/L $, the Hubbard-Stratonovich transformation $ \tr[e^{-\Delta\tau \hat{H}_f}]=\sum_{s^j=\pm 1}^N\tr[e^{-\Delta\tau \hat{H}[s^j]}] $, and then integrates out fermions. We denote $ s^j_i $ as the $ j $-th auxiliary Ising spin on the $ i $-th imaginary time slice. At this stage, the partition function is written purely on the auxiliary Ising spin degrees of freedom $ \sss\equiv\{s_i^j\} $ \cite{Blankenbecler1981,Hirsch1985,White1989}:
\begin{equation}
Z=\sum_{\sss}\det\left[I+\prod_{i=1}^Le^{-\Delta \tau \hat{H}[s_i]}\right]\equiv\sum_{\sss} W[\sss]. \label{eq:weight}
\end{equation}
The Monte Carlo sampling is in the configuration space of $ \mathcal{S} $. The probability $ p $ of accepting a move, for example in the Metropolis-Hastings algorithm, is the weight ratio between two configurations: $ p(\sss_1\to\sss_2)=\min\left(1,W[\sss_2]/W[\sss_1]\right) $. Generally, one must evaluate the determinant in Eq.~\eqref{eq:weight}, which is time-consuming.

The idea of SLMC is to introduce an effective model $ \heff[\sss;\alpha] $ that depends on some trainable parameters $ \alpha $. We would like to optimize $ \alpha $ so that $ W_{\rm eff}[\sss;\alpha]\equiv e^{-\beta\heff[\sss;\alpha]} $ and $ W[\sss] $ are as close as possible. More formally, we would like to minimize the mean squared error (MSE) of the logarithmic weight difference:
\begin{equation}
\min_\alpha \mathbb{E}_{\sss\sim W[\sss]/Z }\left(\ln W_{\rm eff}[\sss;\alpha]-\ln W[\sss] \right)^2.
\label{eq:mse}
\end{equation}
The rationale of minimizing this error will be discussed shortly later. Following the maximum likelihood principle, in practice one could minimize the MSE on a given data set called the training set $ \{\sss,W[\sss]\} $, which is obtained from the Monte Carlo simulation of the original model. This training set is of small size compared with that of the whole configuration space, but is considered typical mimicking the distribution of the original model because it is generated by the importance sampling. Importantly, the training data taken from the Markov chain should be \textit{independently} distributed in order for the maximum likelihood estimation to work well.

One then uses the trained effective model to propose global moves. Starting from a given configuration $ \sss_1 $, one first performs standard Monte Carlo simulations on the effective model $ \sss_1\to\sss_2\to\ldots\to\sss_n $. Configuration $ \sss_n $ is accepted by the original Markov chain with the probability \cite{Liu2017,Liu2017a}
\begin{equation}
p(\sss_1\to\sss_n)=\min\left(1,\frac{W[\sss_n]}{W[\sss_1]}\frac{W_{\rm eff}[\sss_1;\alpha]}{W_{\rm eff}[\sss_n;\alpha]}\right).
\label{eq:acc}
\end{equation}
As proven in Supplemental Material \cite{SM}, the acceptance rate $ \ex{p} $, defined as the expectation of configuration acceptation probability $ p $ defined in Eq.~\eqref{eq:acc}, is directly related to the MSE in Eq.~\eqref{eq:mse} as $\ex{(\ln p)^2} = {\rm MSE}$.  This means MSE serves as a very good estimation of the acceptance rate. Indeed, we will see in the following that these two quantities correspond with each other very well.

The acceleration of SLMC can be analyzed as follows. Denote the computational cost of computing $ W[\sss] $ and $ W_{\rm eff}[\sss;\alpha] $ given $ \sss $ as $ T $ and $ T_{\rm eff} $, and the autocorrelation time of a measurement without SLMC as $ \tau $. Suppose $ T\geq T_{\rm eff} $, one can always to make enough ($ \gtrsim\tau $) updates in the effective model so that the proposed configuration $ \sss_n $ is uncorrelated with $ \sss_1 $. In this way, to obtain two independent configurations without or with SLMC, it takes time $ \mathcal{O}(\tau T) $ and $ \mathcal{O}((\tau T_{\rm eff}+T)/\ex{p}) $. If simulating effective model is efficient $ \tau T_{\rm eff}\ll T $, as demonstrated by cases studied in Ref.~\cite{Liu2017a,Xu2017}, the acceleration is of order $ \ex{p}\tau $.

In principle, there is no limitation on the functional form of $ \heff[\sss;\alpha] $. In the following, we choose to construct $ \heff[\sss;\alpha] $ using neural networks of different architectures. To be concrete, we study the asymmetric Anderson model with a single impurity \cite{Anderson1961}
\begin{align}
\hat{H}=&\hat{H}_0+\hat{H}_1, \\
\hat{H}_0=&\sum_{\mathbf{k}}\varepsilon_\mathbf{k} \hat{c}_\mathbf{k}^\dagger \hat{c}_\mathbf{k}+V\sum_{\mathbf{k}\sigma}(\hat{c}_\mathbf{k}^\dagger \hat{d}_{\sigma}+\mathrm{h.c.})+\mu \hat{n}_d, \\
\hat{H}_1=&U\left(\hat{n}_{d,\uparrow}-\frac{1}{2}\right)\left(\hat{n}_{d,\downarrow}-\frac{1}{2}\right),
\end{align}
where $ \hat{d}_\sigma $ and $ \hat{c}_k $ are the fermion annihilation operator for the impurity and for the conduction electrons respectively. $ \hat{n}_{d,\sigma}\equiv \hat{d}_\sigma^\dagger \hat{d}_\sigma $, $ \hat{n}_d=\sum_{\sigma=\uparrow/\downarrow}\hat{n}_{d,\sigma}$. With different fillings, this model hosts very different low-temperature behaviors identified by the three regimes: local moment, mixed valence and empty orbital \cite{Haldane1978}. The Hubbard-Stratonovich transformation on the impurity site is (up to a constant factor)
\begin{equation}
e^{-\Delta\tau \hat{H}_1}=\frac{1}{2}\sum_{s=\pm 1}e^{\lambda s(\hat{n}_{d,\uparrow}-\hat{n}_{d,\downarrow})},
\label{eq:hst}
\end{equation}
with $ \cosh\lambda=e^{\Delta\tau U/2} $. In total there are $ L $ auxiliary spins, one at each imaginary time slice denoted as a vector $ \mathbf{s}\equiv\sss $. For impurity problems, one may integrate out the continuous band of conducting electrons explicitly and update with Hirsch-Fye algorithm \cite{Hirsch1986,Fye1988}. In the following, the conduction band is assumed to have semicircular density of states $ \rho(\varepsilon)=2\sqrt{1-(\varepsilon/D)^2}/(\pi D) $. The half bandwidth $ D=1 $ is set to be the energy unit.

\begin{figure}[tbp]
\includegraphics[width=\columnwidth]{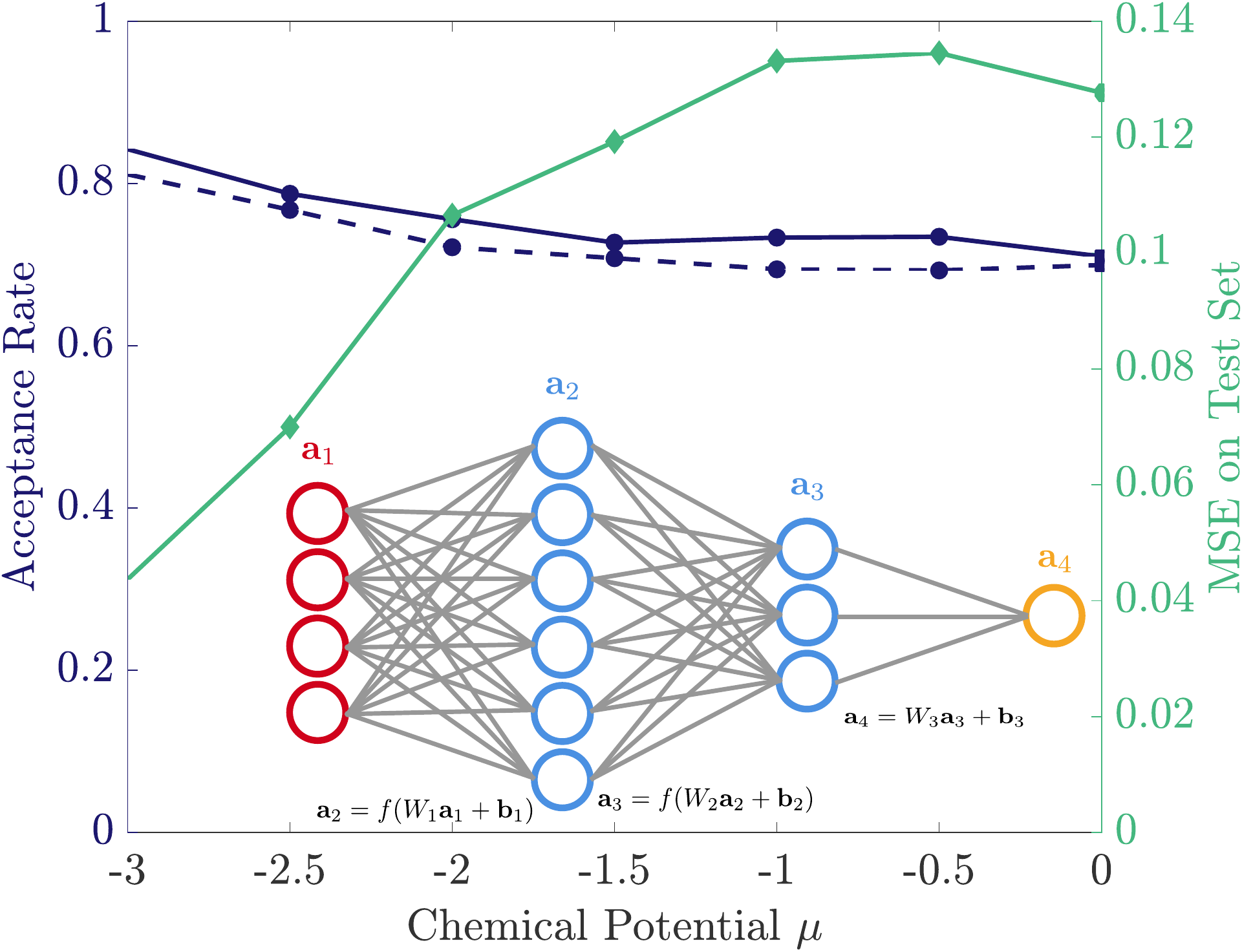}
\caption{The performance of the effective model at different chemical potentials. The dashed blue line is the estimation of acceptance rates through MSE computed in a test data set \cite{SM}: $ e^{-\sqrt{\rm MSE}} $. The discrepancy between the solid and the dashed line is due to the fact that in general $ \ex{p}\neq e^{-\sqrt{\ex{(\ln p)^2}}}=e^{-\sqrt{\rm MSE}} $ and the effective model is biased. Here $\beta=20, U=3.0, V=1.0 $ and $ L=120 $.
The number of neurons in the first and second hidden layers are set as $N_1=100$ and $N_2=50$.
The activation function is the rectified linear unit $ f(x)=\max\{x,0\} $. 
Inset: A schematic show of the fully-connected neural network. The red circles represent neurons in the input layer with $L=4$, and the blue circles represent neurons in two hidden layers with $N_1=6$ and $N_2=3$. The last layer is a linear output layer.
}
\label{fig:fc}
\end{figure}

\textit{Fully-Connected Neural Networks}
To gain some insight on how to design the neural network as the effective model, we first implement the simplest neural network, which consists of several fully-connected layers. Its structure is shown schematically in the inset of  Fig.~\ref{fig:fc}. The effect of $ i $-th fully-connected layer can be summarized as $ \mathbf{a}_{i+1}=f_i(W_i\mathbf{a}_i+\mathbf{b}_i) $, where $ \mathbf{a}_i $ is the input/output vector of the $ i $-th/$(i-1)$-th layer. $ W_i $, $ \mathbf{b}_i $, $ f_i $ are the weight matrix, bias vector, and the nonlinear activation function of such layer. The layer is said to have $ N_i $ neurons when $ W_i $ is of size $ N_i\times N_{i-1} $. This structure as the variational wavefunction in quantum many-body systems has recently been studied extensively  \cite{Carleo2017,Chen2017,Deng2017,Gao2017,Huang2017b,Torlai2017,Hiroki2017,Nomura2017,Clark2017,Glasser2017}.

We take the auxiliary Ising spin as the input vector $ \mathbf{s}\equiv\mathbf{a}_1 $. It is propagated through two nonlinear hidden layers with $ N_1 $ and $ N_2 $ neurons, and then a linear output layer. The output is a number represents the corresponding weight $ \ln W_{\rm eff}[\mathbf{s}] $. We trained the fully-connected neural networks of different architectures and in different physical regimes. The details on the networks and training can be found in Supplemental Material \cite{SM}.

\begin{figure}[tbp]
\includegraphics[width=0.9\columnwidth]{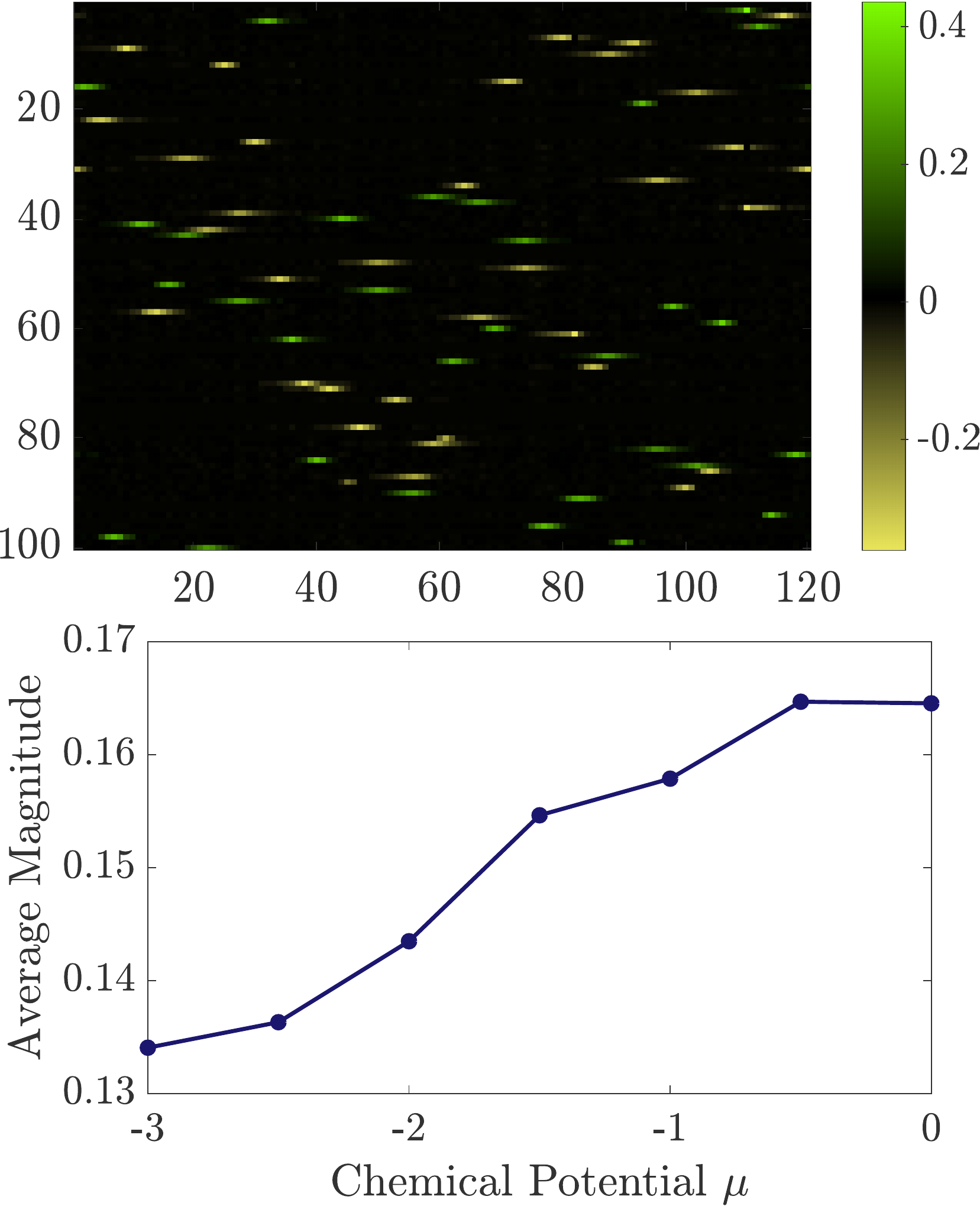}
\caption{Upper: Weight matrix $ W_1 $ taken from the fully-connected neural network of $ \mu=0 $ in Fig.~\ref{fig:fc}. Lower: Average magnitude of nonzero matrix elements in $ W_1 $ of fully-connected neural networks in Fig.~\ref{fig:fc}. The ``nonzero'' matrix element is defined as the element that is greater than 10\% of the maximum element in all 7 weight matrices $ W_1 $ from networks trained at 7 different chemical potentials. Only around 3\% elements are nonzero in these weight matrices. }
\label{fig:weight}
\end{figure}

As shown in Fig.~\ref{fig:fc}, the trained neural network resembles the original model very well. It retains high acceptance rates ($>70\%$) steadily throughout all the parameter regimes. In addition, $ e^{-\sqrt{{\rm MSE}}} $ indeed shares the same trend with the acceptance rate $ \ex{p} $. This suggests that to compare different effective models, one can directly compare the MSE instead of computing the acceptance rate every time.

To extract more features from neural networks, we visualize the weight matrix of the first layers in Fig.~\ref{fig:weight}. The most striking feature is its sparsity. Although the network is fully-connected by construction, most of weight matrix elements vanish after the training, and thus the network is essentially sparsely connected. Clearly, even without any prior knowledge of the system, and using only field configurations and their corresponding energies, the neural network can actually ``learn'' that the correlation between auxiliary spins is short ranged in imaginary time.

Weight matrices in neural networks of different chemical potentials look similar. The main difference lies in the magnitude of the matrix elements. Shown in Fig.~\ref{fig:weight}, the neural network could capture the relative effective interaction strength. When the chemical potential moves away from half-filling, less occupation on the impurity site $ \ex{n_d} $ leads to a weaker coupling between the auxiliary spins and the impurity electrons, according to the Hubbard-Stratonovich transformation Eq.~\eqref{eq:hst}. This further causes the decrease of interaction between the auxiliary spins induced by conducting electrons.

We end this section by briefly discussing the complexity of fully-connected neural networks. The forward propagation of the spin configuration in the fully-connected network involves three matrix-vector multiplications. Each multiplication takes computational cost $ \mathcal{O}(L^2) $, with $ L $ the number of imaginary-time time slices that is usually proportional to $ \beta U $. Thus the running time for a local update in the effective model is $ T_{\rm eff}=\mathcal{O}(L^2) $. In the Hirsch-Fye algorithm, each local update also takes time $ T=\mathcal{O}(L^2) $ \cite{Hirsch1986,Fye1988}. Since $ T_{\rm eff}=T $, SLMC based on fully-connected networks has no advantage in speed over the original Hirsch-Fye algorithm. In the next section, we will show that by taking advantage of the sparse-connection found in the neural network, we can reduce the complexity and make the acceleration possible.

\textit{Exploit the Sparsity}
\begin{figure}[tbp]
\includegraphics[width=0.9\columnwidth]{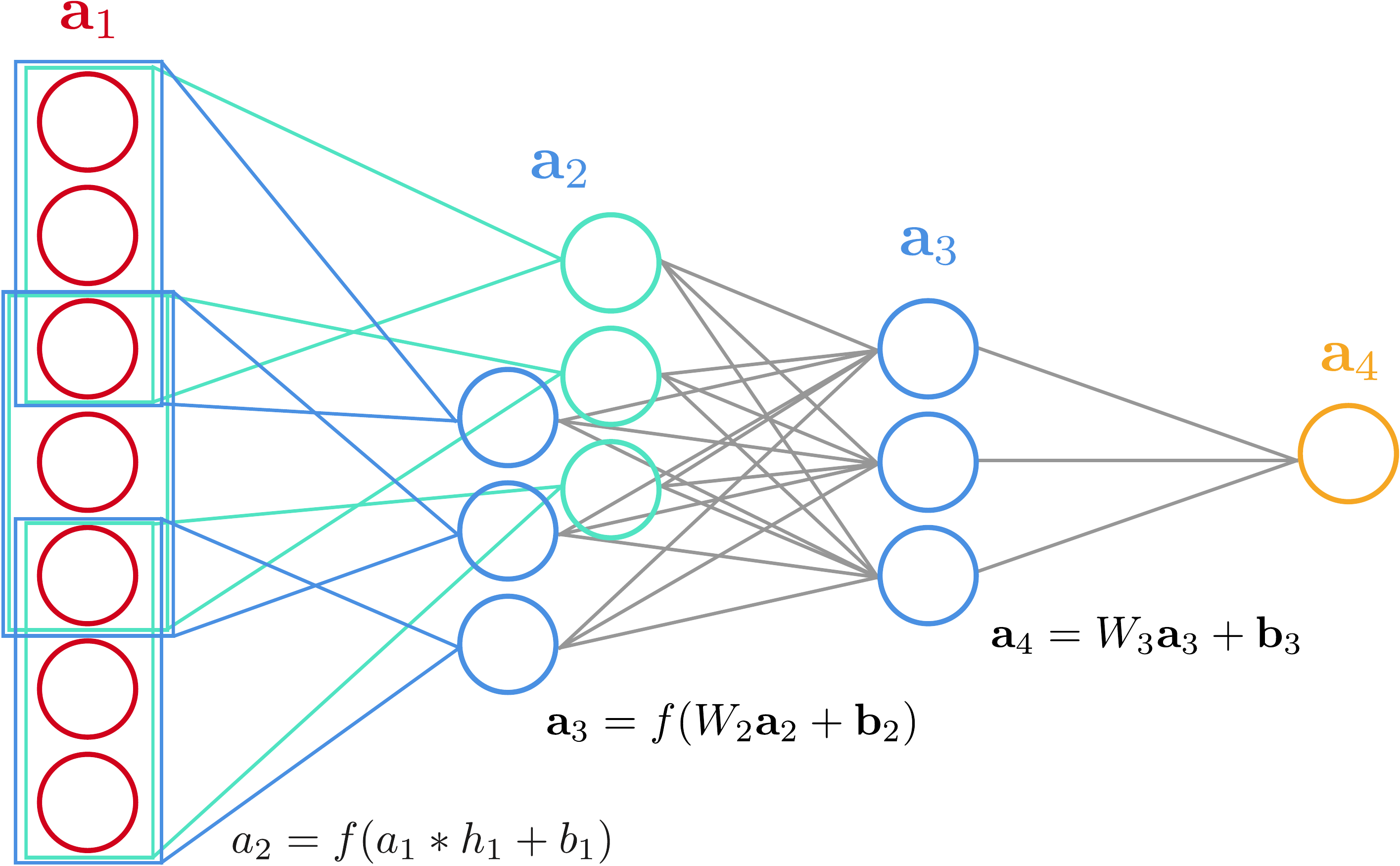}
\caption{The structure of the convolutional neural network. The first layer is a convolutional layer with two kernels (dark and light blue) of size 3. The stride of the sliding inner product is 2. The second layer is a fully-connected layer and the last layer is a linear layer. }
\label{fig:cnnst}
\end{figure}
The sparsity found in the previous section inspires us to design a neural network that is sparsely connected by construction. Moreover, it is known physically that the interaction between the auxiliary spins in imaginary time is translationally invariant, i.e., only depends on $ |\tau_i-\tau_j| $. A neural network that has both properties is known as the ``(1D) convolutional neural network'' \cite{Goodfellow2016}. Instead of doing matrix multiplications as in fully-connected networks, convolutional networks produce their output by sliding inner product denoted by $ * $: $ a_{i+1}=f_i(a_i*h_i+b_i) $ (Fig.~\ref{fig:cnnst}). $ h_i $ and $ b_i $ are called the kernel and the bias of such layer. A detailed mathematical description of such networks can be found in Supplemental Material \cite{SM}.

The key parameters in convolutional neural networks are the number and size of kernels and the stride of sliding inner product. The setup of these parameters can be guided from the fully-connected neural networks: The number and the size of kernels is determined according to the pattern of weight matrices in the fully-connected neural networks, and the stride could be chosen to be half of the kernel size to avoid missing local correlations. For example, for the model whose weight matrix is shown in Fig.~\ref{fig:weight}, we could choose 2 kernels of size 9 for the first convolutional layer with a stride 3. Then several convolutional layers designed in the same spirit are stacked until the size of the output is small enough. Finally, one adds a fully-connected layer to produce the final output. 

Compared with the fully-connected network, the convolutional network has much fewer trainable parameters, which explicitly reduces the redundancy in the parametrization. The fully-connected networks in Fig.~\ref{fig:weight} typically have $ 10^5 $ trainable parameters, while the convolutional networks have only $ 10^2 $ --- smaller by three orders of magnitude.

\begin{figure}[tbp]
\includegraphics[width=.95\columnwidth]{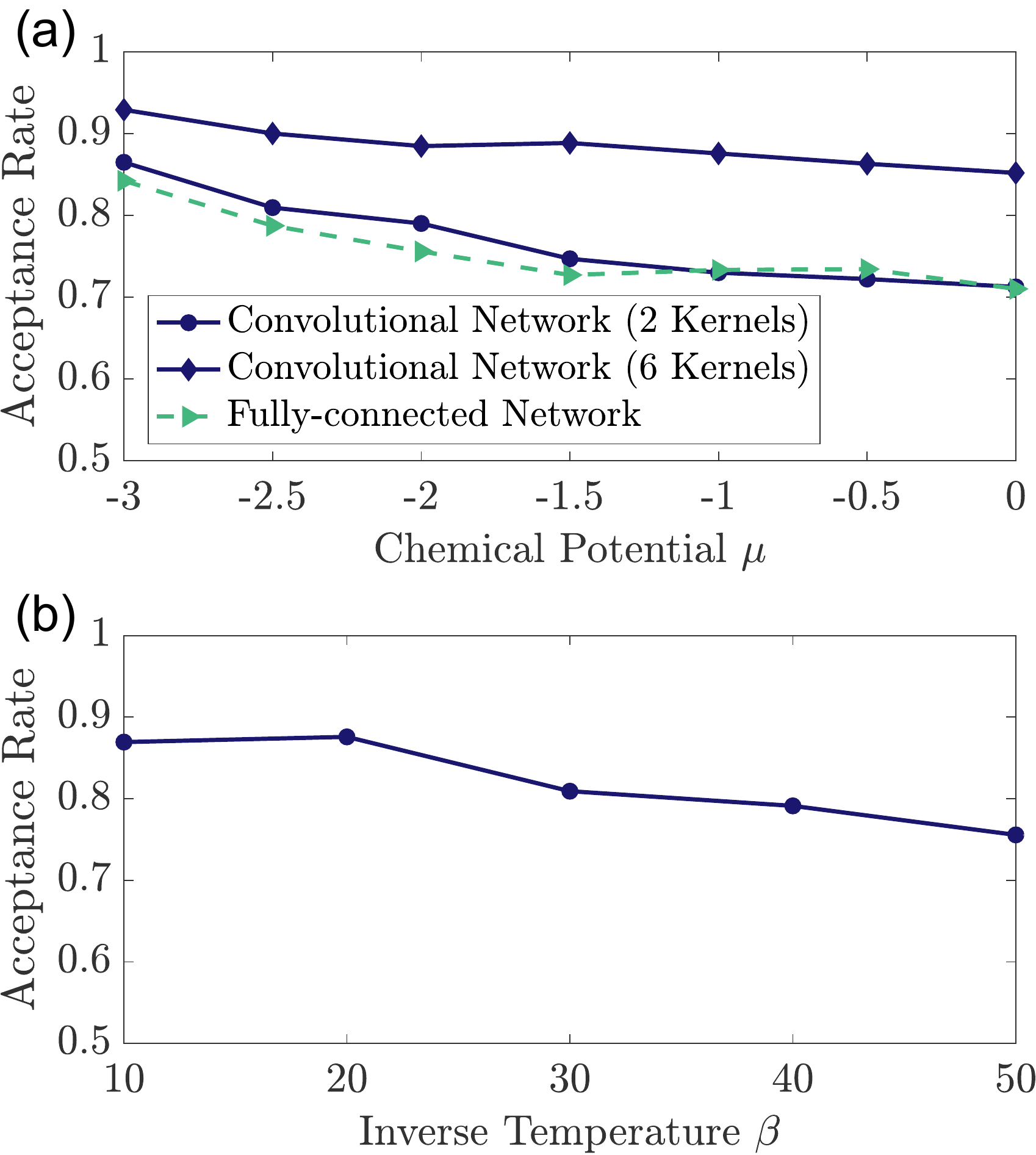}
\caption{(a) The performance of the convolutional network compared with the fully-connected network in Fig.~\ref{fig:fc}. The number of trainable parameters is 211 (2 kernels in the first convolutional layer) or 291 (6 kernels in the first convolutional layer). Here $\beta=20, U=3.0, V=1.0 $ and $ L=120 $. (b) The performance of the convolutional network at different temperatures. Here $U=3.0, \mu=-1.0, V=1.0 $ and $ L=2\beta U $.
The conventional neural network details are described in Supplemental Material \cite{SM}.
}
\label{fig:cnn}
\end{figure}

The performance of convolutional networks are shown in Fig.~\ref{fig:cnn}. The measured fermion imaginary-time Green's function is shown in the Supplemental Material \cite{SM}. Interestingly, the acceptance rate of global moves proposed by convolutional networks are sometimes even higher than those proposed by fully-connected networks. In principle, fully-connected networks with more parameters have greater expressibility. However, for this specific Anderson model, the parameterization of the effective model has a lot of redundancy, as the auxiliary spin interactions are local and are translationally invariant. The convolutional networks reduce this redundancy by construction, and are easier to train potentially due to the smaller parameter space.

The fewer parameters not only make the network easier to train, but also faster to evaluate. Each slide inner product has the computational cost $ \mathcal{O}(L) $. It is important to notice that the strides of the sliding inner product are greater than one so that the dimensions of intermediate outputs keep decreasing. In this way, the number of the convolutional layers is of order $ \mathcal{O}(\ln L) $ because of the large stride. The final fully-connected layer is small. Propagating through such layer only costs a constant computational time that is insensitive to $ L $. To summarize, each local update on the effective model has complexity $ T_{\rm eff}=\mathcal{O}(L\ln L) $,  while that in Hirsch-Fye algorithm is $ T=\mathcal{O}(L^2) $. Since the autocorrelation time for the desired observable is at least of order $ \tau=\Omega(L) $ in order for every auxiliary spin to be updated once, i.e. $\tau T_{\rm eff} \geq  T$, the acceleration with respect to the original Hirsch-Fye algorithm is then $ \tau T/(\tau T_{\rm eff}+T) \approx T/T_{\rm eff}$, of order $ \ex{p}L/\ln L $. It is especially significant for large $ L $. This efficiency allows us to train effective models at very low temperatures very effectively (Fig.~\ref{fig:cnn}(b)), whereas training a fully-connected network is very costly, if possible at all.

\textit{Conclusion}
In this paper, we showed how to integrate neural networks into the framework of SLMC. 
Both the architecture of the networks and the way we design these networks are general and not restricted to impurity models. This work can help design neural networks as effective models in more complicated systems, thereby introducing the state-of-art deep learning hardwares into the field of computational physics.

Particularly for impurity models, we demonstrated that the complexity of the convolutional network for a local update is improved to $ \mathcal{O}(L\ln L) $. 
We note that there exist continuous-time Monte Carlo algorithms that generally outperform the discrete-time Hirsch-Fye algorithm \cite{Gull2007,Gull2011}. Although similar self-learning approaches have already been implemented in these systems \cite{Huang2017a, Nagai2017}, designing an accurate effective model in these continuous-time algorithms is not straightforward as the size of the field configuration keeps changing during the simulation. 
Looking forward, there have already been attempts introducing machine learning into dynamical mean-field theory (DMFT) \cite{Arsenault2014,Arsenault2015}. It will be interesting to accelerate DMFT simulation by integrating SLMC into their impurity solvers \cite{Georges1996}. Moreover, it is worthwhile to develop more advanced network architectures beyond convolutional networks, e.g., networks that are invariant under permutations of the input \cite{Kondor2018}. We leave all these attempts for future work.

\begin{acknowledgments}
H.S. thanks Bo Zeng for helpful discussions. This work is supported by DOE Office of Basic Energy Sciences, Division of Materials Sciences and Engineering under Award DE-SC0010526. J.L. is supported by the start-up funds from HKUST. L.F. is partly supported by the David and Lucile Packard Foundation.
\end{acknowledgments}

\bibliography{HF_Ref}

\begin{thebibliography}{58}%
\makeatletter
\providecommand \@ifxundefined [1]{%
 \@ifx{#1\undefined}
}%
\providecommand \@ifnum [1]{%
 \ifnum #1\expandafter \@firstoftwo
 \else \expandafter \@secondoftwo
 \fi
}%
\providecommand \@ifx [1]{%
 \ifx #1\expandafter \@firstoftwo
 \else \expandafter \@secondoftwo
 \fi
}%
\providecommand \natexlab [1]{#1}%
\providecommand \enquote  [1]{``#1''}%
\providecommand \bibnamefont  [1]{#1}%
\providecommand \bibfnamefont [1]{#1}%
\providecommand \citenamefont [1]{#1}%
\providecommand \href@noop [0]{\@secondoftwo}%
\providecommand \href [0]{\begingroup \@sanitize@url \@href}%
\providecommand \@href[1]{\@@startlink{#1}\@@href}%
\providecommand \@@href[1]{\endgroup#1\@@endlink}%
\providecommand \@sanitize@url [0]{\catcode `\\12\catcode `\$12\catcode
  `\&12\catcode `\#12\catcode `\^12\catcode `\_12\catcode `\%12\relax}%
\providecommand \@@startlink[1]{}%
\providecommand \@@endlink[0]{}%
\providecommand \url  [0]{\begingroup\@sanitize@url \@url }%
\providecommand \@url [1]{\endgroup\@href {#1}{\urlprefix }}%
\providecommand \urlprefix  [0]{URL }%
\providecommand \Eprint [0]{\href }%
\providecommand \doibase [0]{http://dx.doi.org/}%
\providecommand \selectlanguage [0]{\@gobble}%
\providecommand \bibinfo  [0]{\@secondoftwo}%
\providecommand \bibfield  [0]{\@secondoftwo}%
\providecommand \translation [1]{[#1]}%
\providecommand \BibitemOpen [0]{}%
\providecommand \bibitemStop [0]{}%
\providecommand \bibitemNoStop [0]{.\EOS\space}%
\providecommand \EOS [0]{\spacefactor3000\relax}%
\providecommand \BibitemShut  [1]{\csname bibitem#1\endcsname}%
\let\auto@bib@innerbib\@empty
\bibitem [{\citenamefont {Carrasquilla}\ and\ \citenamefont
  {Melko}(2017)}]{Carrasquilla2017}%
  \BibitemOpen
  \bibfield  {author} {\bibinfo {author} {\bibfnamefont {J.}~\bibnamefont
  {Carrasquilla}}\ and\ \bibinfo {author} {\bibfnamefont {R.~G.}\ \bibnamefont
  {Melko}},\ }\href {\doibase 10.1038/nphys4035} {\bibfield  {journal}
  {\bibinfo  {journal} {Nat. Phys.}\ }\textbf {\bibinfo {volume} {13}},\
  \bibinfo {pages} {431} (\bibinfo {year} {2017})}\BibitemShut {NoStop}%
\bibitem [{\citenamefont {Wang}(2016)}]{Wang2016a}%
  \BibitemOpen
  \bibfield  {author} {\bibinfo {author} {\bibfnamefont {L.}~\bibnamefont
  {Wang}},\ }\href {\doibase 10.1103/PhysRevB.94.195105} {\bibfield  {journal}
  {\bibinfo  {journal} {Phys. Rev. B}\ }\textbf {\bibinfo {volume} {94}},\
  \bibinfo {pages} {195105} (\bibinfo {year} {2016})}\BibitemShut {NoStop}%
\bibitem [{\citenamefont {Tanaka}\ and\ \citenamefont
  {Tomiya}(2017)}]{Tanaka2017a}%
  \BibitemOpen
  \bibfield  {author} {\bibinfo {author} {\bibfnamefont {A.}~\bibnamefont
  {Tanaka}}\ and\ \bibinfo {author} {\bibfnamefont {A.}~\bibnamefont
  {Tomiya}},\ }\href {\doibase 10.7566/JPSJ.86.063001} {\bibfield  {journal}
  {\bibinfo  {journal} {J. Phys. Soc. Japan}\ }\textbf {\bibinfo {volume}
  {86}},\ \bibinfo {pages} {063001} (\bibinfo {year} {2017})}\BibitemShut
  {NoStop}%
\bibitem [{\citenamefont {Ohtsuki}\ and\ \citenamefont
  {Ohtsuki}(2016)}]{Ohtsuki2017}%
  \BibitemOpen
  \bibfield  {author} {\bibinfo {author} {\bibfnamefont {T.}~\bibnamefont
  {Ohtsuki}}\ and\ \bibinfo {author} {\bibfnamefont {T.}~\bibnamefont
  {Ohtsuki}},\ }\href {\doibase 10.7566/JPSJ.85.123706} {\bibfield  {journal}
  {\bibinfo  {journal} {J. Phys. Soc. Japan}\ }\textbf {\bibinfo {volume}
  {85}},\ \bibinfo {pages} {123706} (\bibinfo {year} {2016})}\BibitemShut
  {NoStop}%
\bibitem [{\citenamefont {van Nieuwenburg}\ \emph {et~al.}(2017)\citenamefont
  {van Nieuwenburg}, \citenamefont {Liu},\ and\ \citenamefont
  {Huber}}]{VanNieuwenburg2017}%
  \BibitemOpen
  \bibfield  {author} {\bibinfo {author} {\bibfnamefont {E.~P.~L.}\
  \bibnamefont {van Nieuwenburg}}, \bibinfo {author} {\bibfnamefont {Y.-H.}\
  \bibnamefont {Liu}}, \ and\ \bibinfo {author} {\bibfnamefont {S.~D.}\
  \bibnamefont {Huber}},\ }\href {\doibase 10.1038/nphys4037} {\bibfield
  {journal} {\bibinfo  {journal} {Nat. Phys.}\ }\textbf {\bibinfo {volume}
  {13}},\ \bibinfo {pages} {435} (\bibinfo {year} {2017})}\BibitemShut
  {NoStop}%
\bibitem [{\citenamefont {Zhang}\ and\ \citenamefont {Kim}(2017)}]{Zhang2017a}%
  \BibitemOpen
  \bibfield  {author} {\bibinfo {author} {\bibfnamefont {Y.}~\bibnamefont
  {Zhang}}\ and\ \bibinfo {author} {\bibfnamefont {E.-A.}\ \bibnamefont
  {Kim}},\ }\href {\doibase 10.1103/PhysRevLett.118.216401} {\bibfield
  {journal} {\bibinfo  {journal} {Phys. Rev. Lett.}\ }\textbf {\bibinfo
  {volume} {118}},\ \bibinfo {pages} {216401} (\bibinfo {year}
  {2017})}\BibitemShut {NoStop}%
\bibitem [{\citenamefont {Mills}\ \emph {et~al.}(2017)\citenamefont {Mills},
  \citenamefont {Spanner},\ and\ \citenamefont {Tamblyn}}]{Mills2017a}%
  \BibitemOpen
  \bibfield  {author} {\bibinfo {author} {\bibfnamefont {K.}~\bibnamefont
  {Mills}}, \bibinfo {author} {\bibfnamefont {M.}~\bibnamefont {Spanner}}, \
  and\ \bibinfo {author} {\bibfnamefont {I.}~\bibnamefont {Tamblyn}},\ }\href
  {\doibase 10.1103/PhysRevA.96.042113} {\bibfield  {journal} {\bibinfo
  {journal} {Phys. Rev. A}\ }\textbf {\bibinfo {volume} {96}},\ \bibinfo
  {pages} {042113} (\bibinfo {year} {2017})}\BibitemShut {NoStop}%
\bibitem [{\citenamefont {Wetzel}(2017)}]{Wetzel2017}%
  \BibitemOpen
  \bibfield  {author} {\bibinfo {author} {\bibfnamefont {S.~J.}\ \bibnamefont
  {Wetzel}},\ }\href {\doibase 10.1103/PhysRevE.96.022140} {\bibfield
  {journal} {\bibinfo  {journal} {Phys. Rev. E}\ }\textbf {\bibinfo {volume}
  {96}},\ \bibinfo {pages} {022140} (\bibinfo {year} {2017})}\BibitemShut
  {NoStop}%
\bibitem [{\citenamefont {Hu}\ \emph {et~al.}(2017)\citenamefont {Hu},
  \citenamefont {Singh},\ and\ \citenamefont {Scalettar}}]{Hu2017a}%
  \BibitemOpen
  \bibfield  {author} {\bibinfo {author} {\bibfnamefont {W.}~\bibnamefont
  {Hu}}, \bibinfo {author} {\bibfnamefont {R.~R.~P.}\ \bibnamefont {Singh}}, \
  and\ \bibinfo {author} {\bibfnamefont {R.~T.}\ \bibnamefont {Scalettar}},\
  }\href {\doibase 10.1103/PhysRevE.95.062122} {\bibfield  {journal} {\bibinfo
  {journal} {Phys. Rev. E}\ }\textbf {\bibinfo {volume} {95}},\ \bibinfo
  {pages} {062122} (\bibinfo {year} {2017})}\BibitemShut {NoStop}%
\bibitem [{\citenamefont {Schindler}\ \emph {et~al.}(2017)\citenamefont
  {Schindler}, \citenamefont {Regnault},\ and\ \citenamefont
  {Neupert}}]{Schindler2017}%
  \BibitemOpen
  \bibfield  {author} {\bibinfo {author} {\bibfnamefont {F.}~\bibnamefont
  {Schindler}}, \bibinfo {author} {\bibfnamefont {N.}~\bibnamefont {Regnault}},
  \ and\ \bibinfo {author} {\bibfnamefont {T.}~\bibnamefont {Neupert}},\ }\href
  {\doibase 10.1103/PhysRevB.95.245134} {\bibfield  {journal} {\bibinfo
  {journal} {Phys. Rev. B}\ }\textbf {\bibinfo {volume} {95}},\ \bibinfo
  {pages} {245134} (\bibinfo {year} {2017})}\BibitemShut {NoStop}%
\bibitem [{\citenamefont {Lu}\ \emph {et~al.}()\citenamefont {Lu},
  \citenamefont {Huang}, \citenamefont {Li}, \citenamefont {Li}, \citenamefont
  {Chen}, \citenamefont {Lu}, \citenamefont {Ji}, \citenamefont {Shen},
  \citenamefont {Zhou},\ and\ \citenamefont {Zeng}}]{Lu2017}%
  \BibitemOpen
  \bibfield  {author} {\bibinfo {author} {\bibfnamefont {S.}~\bibnamefont
  {Lu}}, \bibinfo {author} {\bibfnamefont {S.}~\bibnamefont {Huang}}, \bibinfo
  {author} {\bibfnamefont {K.}~\bibnamefont {Li}}, \bibinfo {author}
  {\bibfnamefont {J.}~\bibnamefont {Li}}, \bibinfo {author} {\bibfnamefont
  {J.}~\bibnamefont {Chen}}, \bibinfo {author} {\bibfnamefont {D.}~\bibnamefont
  {Lu}}, \bibinfo {author} {\bibfnamefont {Z.}~\bibnamefont {Ji}}, \bibinfo
  {author} {\bibfnamefont {Y.}~\bibnamefont {Shen}}, \bibinfo {author}
  {\bibfnamefont {D.}~\bibnamefont {Zhou}}, \ and\ \bibinfo {author}
  {\bibfnamefont {B.}~\bibnamefont {Zeng}},\ }\href
  {http://arxiv.org/abs/1705.01523} {\ }\Eprint
  {http://arxiv.org/abs/1705.01523} {arXiv:1705.01523} \BibitemShut {NoStop}%
\bibitem [{\citenamefont {Cristoforetti}\ \emph {et~al.}()\citenamefont
  {Cristoforetti}, \citenamefont {Jurman}, \citenamefont {Nardelli},\ and\
  \citenamefont {Furlanello}}]{Cristoforetti2017}%
  \BibitemOpen
  \bibfield  {author} {\bibinfo {author} {\bibfnamefont {M.}~\bibnamefont
  {Cristoforetti}}, \bibinfo {author} {\bibfnamefont {G.}~\bibnamefont
  {Jurman}}, \bibinfo {author} {\bibfnamefont {A.~I.}\ \bibnamefont
  {Nardelli}}, \ and\ \bibinfo {author} {\bibfnamefont {C.}~\bibnamefont
  {Furlanello}},\ }\href {http://arxiv.org/abs/1705.09524} {\ }\Eprint
  {http://arxiv.org/abs/1705.09524} {arXiv:1705.09524} \BibitemShut {NoStop}%
\bibitem [{\citenamefont {Wang}\ and\ \citenamefont {Zhai}(2017)}]{Wang2017}%
  \BibitemOpen
  \bibfield  {author} {\bibinfo {author} {\bibfnamefont {C.}~\bibnamefont
  {Wang}}\ and\ \bibinfo {author} {\bibfnamefont {H.}~\bibnamefont {Zhai}},\
  }\href {\doibase 10.1103/PhysRevB.96.144432} {\bibfield  {journal} {\bibinfo
  {journal} {Phys. Rev. B}\ }\textbf {\bibinfo {volume} {96}},\ \bibinfo
  {pages} {144432} (\bibinfo {year} {2017})}\BibitemShut {NoStop}%
\bibitem [{\citenamefont {Zhang}\ \emph {et~al.}(2018)\citenamefont {Zhang},
  \citenamefont {Shen},\ and\ \citenamefont {Zhai}}]{Zhang2017}%
  \BibitemOpen
  \bibfield  {author} {\bibinfo {author} {\bibfnamefont {P.}~\bibnamefont
  {Zhang}}, \bibinfo {author} {\bibfnamefont {H.}~\bibnamefont {Shen}}, \ and\
  \bibinfo {author} {\bibfnamefont {H.}~\bibnamefont {Zhai}},\ }\href {\doibase
  10.1103/PhysRevLett.120.066401} {\bibfield  {journal} {\bibinfo  {journal}
  {Phys. Rev. Lett.}\ }\textbf {\bibinfo {volume} {120}},\ \bibinfo {pages}
  {066401} (\bibinfo {year} {2018})}\BibitemShut {NoStop}%
\bibitem [{\citenamefont {Rao}\ \emph {et~al.}(2018)\citenamefont {Rao},
  \citenamefont {Li}, \citenamefont {Zhu}, \citenamefont {Luo},\ and\
  \citenamefont {Wan}}]{Rao2017}%
  \BibitemOpen
  \bibfield  {author} {\bibinfo {author} {\bibfnamefont {W.-J.}\ \bibnamefont
  {Rao}}, \bibinfo {author} {\bibfnamefont {Z.}~\bibnamefont {Li}}, \bibinfo
  {author} {\bibfnamefont {Q.}~\bibnamefont {Zhu}}, \bibinfo {author}
  {\bibfnamefont {M.}~\bibnamefont {Luo}}, \ and\ \bibinfo {author}
  {\bibfnamefont {X.}~\bibnamefont {Wan}},\ }\href {\doibase
  10.1103/PhysRevB.97.094207} {\bibfield  {journal} {\bibinfo  {journal} {Phys.
  Rev. B}\ }\textbf {\bibinfo {volume} {97}},\ \bibinfo {pages} {094207}
  (\bibinfo {year} {2018})}\BibitemShut {NoStop}%
\bibitem [{\citenamefont {Yoshioka}\ \emph {et~al.}(2018)\citenamefont
  {Yoshioka}, \citenamefont {Akagi},\ and\ \citenamefont
  {Katsura}}]{Yoshioka2017}%
  \BibitemOpen
  \bibfield  {author} {\bibinfo {author} {\bibfnamefont {N.}~\bibnamefont
  {Yoshioka}}, \bibinfo {author} {\bibfnamefont {Y.}~\bibnamefont {Akagi}}, \
  and\ \bibinfo {author} {\bibfnamefont {H.}~\bibnamefont {Katsura}},\ }\href
  {\doibase 10.1103/PhysRevB.97.205110} {\bibfield  {journal} {\bibinfo
  {journal} {Phys. Rev. B}\ }\textbf {\bibinfo {volume} {97}},\ \bibinfo
  {pages} {205110} (\bibinfo {year} {2018})}\BibitemShut {NoStop}%
\bibitem [{\citenamefont {Huembeli}\ \emph {et~al.}(2018)\citenamefont
  {Huembeli}, \citenamefont {Dauphin},\ and\ \citenamefont
  {Wittek}}]{Huembeli2017}%
  \BibitemOpen
  \bibfield  {author} {\bibinfo {author} {\bibfnamefont {P.}~\bibnamefont
  {Huembeli}}, \bibinfo {author} {\bibfnamefont {A.}~\bibnamefont {Dauphin}}, \
  and\ \bibinfo {author} {\bibfnamefont {P.}~\bibnamefont {Wittek}},\ }\href
  {\doibase 10.1103/PhysRevB.97.134109} {\bibfield  {journal} {\bibinfo
  {journal} {Phys. Rev. B}\ }\textbf {\bibinfo {volume} {97}},\ \bibinfo
  {pages} {134109} (\bibinfo {year} {2018})}\BibitemShut {NoStop}%
\bibitem [{\citenamefont {Liu}\ \emph {et~al.}(2017{\natexlab{a}})\citenamefont
  {Liu}, \citenamefont {Qi}, \citenamefont {Meng},\ and\ \citenamefont
  {Fu}}]{Liu2017}%
  \BibitemOpen
  \bibfield  {author} {\bibinfo {author} {\bibfnamefont {J.}~\bibnamefont
  {Liu}}, \bibinfo {author} {\bibfnamefont {Y.}~\bibnamefont {Qi}}, \bibinfo
  {author} {\bibfnamefont {Z.~Y.}\ \bibnamefont {Meng}}, \ and\ \bibinfo
  {author} {\bibfnamefont {L.}~\bibnamefont {Fu}},\ }\href {\doibase
  10.1103/PhysRevB.95.041101} {\bibfield  {journal} {\bibinfo  {journal} {Phys.
  Rev. B}\ }\textbf {\bibinfo {volume} {95}},\ \bibinfo {pages} {041101}
  (\bibinfo {year} {2017}{\natexlab{a}})}\BibitemShut {NoStop}%
\bibitem [{\citenamefont {Huang}\ and\ \citenamefont {Wang}(2017)}]{Huang2017}%
  \BibitemOpen
  \bibfield  {author} {\bibinfo {author} {\bibfnamefont {L.}~\bibnamefont
  {Huang}}\ and\ \bibinfo {author} {\bibfnamefont {L.}~\bibnamefont {Wang}},\
  }\href {\doibase 10.1103/PhysRevB.95.035105} {\bibfield  {journal} {\bibinfo
  {journal} {Phys. Rev. B}\ }\textbf {\bibinfo {volume} {95}},\ \bibinfo
  {pages} {035105} (\bibinfo {year} {2017})}\BibitemShut {NoStop}%
\bibitem [{\citenamefont {Liu}\ \emph {et~al.}(2017{\natexlab{b}})\citenamefont
  {Liu}, \citenamefont {Shen}, \citenamefont {Qi}, \citenamefont {Meng},\ and\
  \citenamefont {Fu}}]{Liu2017a}%
  \BibitemOpen
  \bibfield  {author} {\bibinfo {author} {\bibfnamefont {J.}~\bibnamefont
  {Liu}}, \bibinfo {author} {\bibfnamefont {H.}~\bibnamefont {Shen}}, \bibinfo
  {author} {\bibfnamefont {Y.}~\bibnamefont {Qi}}, \bibinfo {author}
  {\bibfnamefont {Z.~Y.}\ \bibnamefont {Meng}}, \ and\ \bibinfo {author}
  {\bibfnamefont {L.}~\bibnamefont {Fu}},\ }\href {\doibase
  10.1103/PhysRevB.95.241104} {\bibfield  {journal} {\bibinfo  {journal} {Phys.
  Rev. B}\ }\textbf {\bibinfo {volume} {95}},\ \bibinfo {pages} {241104}
  (\bibinfo {year} {2017}{\natexlab{b}})}\BibitemShut {NoStop}%
\bibitem [{\citenamefont {Xu}\ \emph {et~al.}(2017)\citenamefont {Xu},
  \citenamefont {Qi}, \citenamefont {Liu}, \citenamefont {Fu},\ and\
  \citenamefont {Meng}}]{Xu2017}%
  \BibitemOpen
  \bibfield  {author} {\bibinfo {author} {\bibfnamefont {X.~Y.}\ \bibnamefont
  {Xu}}, \bibinfo {author} {\bibfnamefont {Y.}~\bibnamefont {Qi}}, \bibinfo
  {author} {\bibfnamefont {J.}~\bibnamefont {Liu}}, \bibinfo {author}
  {\bibfnamefont {L.}~\bibnamefont {Fu}}, \ and\ \bibinfo {author}
  {\bibfnamefont {Z.~Y.}\ \bibnamefont {Meng}},\ }\href {\doibase
  10.1103/PhysRevB.96.041119} {\bibfield  {journal} {\bibinfo  {journal} {Phys.
  Rev. B}\ }\textbf {\bibinfo {volume} {96}},\ \bibinfo {pages} {041119}
  (\bibinfo {year} {2017})}\BibitemShut {NoStop}%
\bibitem [{\citenamefont {Huang}\ \emph {et~al.}(2017)\citenamefont {Huang},
  \citenamefont {Yang},\ and\ \citenamefont {Wang}}]{Huang2017a}%
  \BibitemOpen
  \bibfield  {author} {\bibinfo {author} {\bibfnamefont {L.}~\bibnamefont
  {Huang}}, \bibinfo {author} {\bibfnamefont {Y.-f.}\ \bibnamefont {Yang}}, \
  and\ \bibinfo {author} {\bibfnamefont {L.}~\bibnamefont {Wang}},\ }\href
  {\doibase 10.1103/PhysRevE.95.031301} {\bibfield  {journal} {\bibinfo
  {journal} {Phys. Rev. E}\ }\textbf {\bibinfo {volume} {95}},\ \bibinfo
  {pages} {031301} (\bibinfo {year} {2017})}\BibitemShut {NoStop}%
\bibitem [{\citenamefont {Nagai}\ \emph {et~al.}(2017)\citenamefont {Nagai},
  \citenamefont {Shen}, \citenamefont {Qi}, \citenamefont {Liu},\ and\
  \citenamefont {Fu}}]{Nagai2017}%
  \BibitemOpen
  \bibfield  {author} {\bibinfo {author} {\bibfnamefont {Y.}~\bibnamefont
  {Nagai}}, \bibinfo {author} {\bibfnamefont {H.}~\bibnamefont {Shen}},
  \bibinfo {author} {\bibfnamefont {Y.}~\bibnamefont {Qi}}, \bibinfo {author}
  {\bibfnamefont {J.}~\bibnamefont {Liu}}, \ and\ \bibinfo {author}
  {\bibfnamefont {L.}~\bibnamefont {Fu}},\ }\href {\doibase
  10.1103/PhysRevB.96.161102} {\bibfield  {journal} {\bibinfo  {journal} {Phys.
  Rev. B}\ }\textbf {\bibinfo {volume} {96}},\ \bibinfo {pages} {161102}
  (\bibinfo {year} {2017})}\BibitemShut {NoStop}%
\bibitem [{\citenamefont {Tanaka}\ and\ \citenamefont {Tomiya}()}]{Tanaka2017}%
  \BibitemOpen
  \bibfield  {author} {\bibinfo {author} {\bibfnamefont {A.}~\bibnamefont
  {Tanaka}}\ and\ \bibinfo {author} {\bibfnamefont {A.}~\bibnamefont
  {Tomiya}},\ }\href {https://arxiv.org/abs/1712.03893} {\ }\Eprint
  {http://arxiv.org/abs/1712.03893} {arXiv:1712.03893} \BibitemShut {NoStop}%
\bibitem [{\citenamefont {Liu}\ \emph {et~al.}()\citenamefont {Liu},
  \citenamefont {Xu}, \citenamefont {Qi}, \citenamefont {Sun},\ and\
  \citenamefont {Meng}}]{Liu2017b}%
  \BibitemOpen
  \bibfield  {author} {\bibinfo {author} {\bibfnamefont {Z.~H.}\ \bibnamefont
  {Liu}}, \bibinfo {author} {\bibfnamefont {X.~Y.}\ \bibnamefont {Xu}},
  \bibinfo {author} {\bibfnamefont {Y.}~\bibnamefont {Qi}}, \bibinfo {author}
  {\bibfnamefont {K.}~\bibnamefont {Sun}}, \ and\ \bibinfo {author}
  {\bibfnamefont {Z.~Y.}\ \bibnamefont {Meng}},\ }\href
  {http://arxiv.org/abs/1706.10004} {\ }\Eprint
  {http://arxiv.org/abs/1706.10004} {arXiv:1706.10004} \BibitemShut {NoStop}%
\bibitem [{\citenamefont {Baroni}\ and\ \citenamefont
  {Moroni}(1999)}]{PhysRevLett.82.4745}%
  \BibitemOpen
  \bibfield  {author} {\bibinfo {author} {\bibfnamefont {S.}~\bibnamefont
  {Baroni}}\ and\ \bibinfo {author} {\bibfnamefont {S.}~\bibnamefont
  {Moroni}},\ }\href {\doibase 10.1103/PhysRevLett.82.4745} {\bibfield
  {journal} {\bibinfo  {journal} {Phys. Rev. Lett.}\ }\textbf {\bibinfo
  {volume} {82}},\ \bibinfo {pages} {4745} (\bibinfo {year}
  {1999})}\BibitemShut {NoStop}%
\bibitem [{\citenamefont {Pierleoni}\ and\ \citenamefont
  {Ceperley}(2005)}]{doi:10.1002/cphc.200400587}%
  \BibitemOpen
  \bibfield  {author} {\bibinfo {author} {\bibfnamefont {C.}~\bibnamefont
  {Pierleoni}}\ and\ \bibinfo {author} {\bibfnamefont {D.~M.}\ \bibnamefont
  {Ceperley}},\ }\href {\doibase 10.1002/cphc.200400587} {\bibfield  {journal}
  {\bibinfo  {journal} {ChemPhysChem}\ }\textbf {\bibinfo {volume} {6}},\
  \bibinfo {pages} {1872} (\bibinfo {year} {2005})}\BibitemShut {NoStop}%
\bibitem [{\citenamefont {Sylju\aa{}sen}\ and\ \citenamefont
  {Sandvik}(2002)}]{PhysRevE.66.046701}%
  \BibitemOpen
  \bibfield  {author} {\bibinfo {author} {\bibfnamefont {O.~F.}\ \bibnamefont
  {Sylju\aa{}sen}}\ and\ \bibinfo {author} {\bibfnamefont {A.~W.}\ \bibnamefont
  {Sandvik}},\ }\href {\doibase 10.1103/PhysRevE.66.046701} {\bibfield
  {journal} {\bibinfo  {journal} {Phys. Rev. E}\ }\textbf {\bibinfo {volume}
  {66}},\ \bibinfo {pages} {046701} (\bibinfo {year} {2002})}\BibitemShut
  {NoStop}%
\bibitem [{\citenamefont {Rousseau}(2008)}]{PhysRevE.78.056707}%
  \BibitemOpen
  \bibfield  {author} {\bibinfo {author} {\bibfnamefont {V.~G.}\ \bibnamefont
  {Rousseau}},\ }\href {\doibase 10.1103/PhysRevE.78.056707} {\bibfield
  {journal} {\bibinfo  {journal} {Phys. Rev. E}\ }\textbf {\bibinfo {volume}
  {78}},\ \bibinfo {pages} {056707} (\bibinfo {year} {2008})}\BibitemShut
  {NoStop}%
\bibitem [{\citenamefont {Carleo}\ \emph {et~al.}(2010)\citenamefont {Carleo},
  \citenamefont {Becca}, \citenamefont {Moroni},\ and\ \citenamefont
  {Baroni}}]{PhysRevE.82.046710}%
  \BibitemOpen
  \bibfield  {author} {\bibinfo {author} {\bibfnamefont {G.}~\bibnamefont
  {Carleo}}, \bibinfo {author} {\bibfnamefont {F.}~\bibnamefont {Becca}},
  \bibinfo {author} {\bibfnamefont {S.}~\bibnamefont {Moroni}}, \ and\ \bibinfo
  {author} {\bibfnamefont {S.}~\bibnamefont {Baroni}},\ }\href {\doibase
  10.1103/PhysRevE.82.046710} {\bibfield  {journal} {\bibinfo  {journal} {Phys.
  Rev. E}\ }\textbf {\bibinfo {volume} {82}},\ \bibinfo {pages} {046710}
  (\bibinfo {year} {2010})}\BibitemShut {NoStop}%
\bibitem [{\citenamefont {Cybenko}(1989)}]{Cybenko1989}%
  \BibitemOpen
  \bibfield  {author} {\bibinfo {author} {\bibfnamefont {G.}~\bibnamefont
  {Cybenko}},\ }\href {\doibase 10.1007/BF02551274} {\bibfield  {journal}
  {\bibinfo  {journal} {Math. Control. Signals, Syst.}\ }\textbf {\bibinfo
  {volume} {2}},\ \bibinfo {pages} {303} (\bibinfo {year} {1989})}\BibitemShut
  {NoStop}%
\bibitem [{\citenamefont {Hornik}(1991)}]{Hornik1991}%
  \BibitemOpen
  \bibfield  {author} {\bibinfo {author} {\bibfnamefont {K.}~\bibnamefont
  {Hornik}},\ }\href {\doibase 10.1016/0893-6080(91)90009-T} {\bibfield
  {journal} {\bibinfo  {journal} {Neural Networks}\ }\textbf {\bibinfo {volume}
  {4}},\ \bibinfo {pages} {251} (\bibinfo {year} {1991})}\BibitemShut {NoStop}%
\bibitem [{\citenamefont {Goodfellow}\ \emph {et~al.}(2016)\citenamefont
  {Goodfellow}, \citenamefont {Bengio},\ and\ \citenamefont
  {Courville}}]{Goodfellow2016}%
  \BibitemOpen
  \bibfield  {author} {\bibinfo {author} {\bibfnamefont {I.}~\bibnamefont
  {Goodfellow}}, \bibinfo {author} {\bibfnamefont {Y.}~\bibnamefont {Bengio}},
  \ and\ \bibinfo {author} {\bibfnamefont {A.}~\bibnamefont {Courville}},\
  }\href@noop {} {\emph {\bibinfo {title} {Deep Learning}}}\ (\bibinfo
  {publisher} {MIT Press},\ \bibinfo {year} {2016})\ \bibinfo {note}
  {\url{http://www.deeplearningbook.org}}\BibitemShut {NoStop}%
\bibitem [{\citenamefont {Jouppi}\ \emph {et~al.}(2017)\citenamefont {Jouppi},
  \citenamefont {Young}, \citenamefont {Patil}, \citenamefont {Patterson},
  \citenamefont {Agrawal}, \citenamefont {Bajwa}, \citenamefont {Bates},
  \citenamefont {Bhatia}, \citenamefont {Boden}, \citenamefont {Borchers},
  \citenamefont {Boyle}, \citenamefont {Cantin}, \citenamefont {Chao},
  \citenamefont {Clark}, \citenamefont {Coriell}, \citenamefont {Daley},
  \citenamefont {Dau}, \citenamefont {Dean}, \citenamefont {Gelb},
  \citenamefont {Ghaemmaghami}, \citenamefont {Gottipati}, \citenamefont
  {Gulland}, \citenamefont {Hagmann}, \citenamefont {Ho}, \citenamefont
  {Hogberg}, \citenamefont {Hu}, \citenamefont {Hundt}, \citenamefont {Hurt},
  \citenamefont {Ibarz}, \citenamefont {Jaffey}, \citenamefont {Jaworski},
  \citenamefont {Kaplan}, \citenamefont {Khaitan}, \citenamefont {Killebrew},
  \citenamefont {Koch}, \citenamefont {Kumar}, \citenamefont {Lacy},
  \citenamefont {Laudon}, \citenamefont {Law}, \citenamefont {Le},
  \citenamefont {Leary}, \citenamefont {Liu}, \citenamefont {Lucke},
  \citenamefont {Lundin}, \citenamefont {MacKean}, \citenamefont {Maggiore},
  \citenamefont {Mahony}, \citenamefont {Miller}, \citenamefont {Nagarajan},
  \citenamefont {Narayanaswami}, \citenamefont {Ni}, \citenamefont {Nix},
  \citenamefont {Norrie}, \citenamefont {Omernick}, \citenamefont {Penukonda},
  \citenamefont {Phelps}, \citenamefont {Ross}, \citenamefont {Ross},
  \citenamefont {Salek}, \citenamefont {Samadiani}, \citenamefont {Severn},
  \citenamefont {Sizikov}, \citenamefont {Snelham}, \citenamefont {Souter},
  \citenamefont {Steinberg}, \citenamefont {Swing}, \citenamefont {Tan},
  \citenamefont {Thorson}, \citenamefont {Tian}, \citenamefont {Toma},
  \citenamefont {Tuttle}, \citenamefont {Vasudevan}, \citenamefont {Walter},
  \citenamefont {Wang}, \citenamefont {Wilcox},\ and\ \citenamefont
  {Yoon}}]{Jouppi2017}%
  \BibitemOpen
  \bibfield  {author} {\bibinfo {author} {\bibfnamefont {N.~P.}\ \bibnamefont
  {Jouppi}}, \bibinfo {author} {\bibfnamefont {C.}~\bibnamefont {Young}},
  \bibinfo {author} {\bibfnamefont {N.}~\bibnamefont {Patil}}, \bibinfo
  {author} {\bibfnamefont {D.}~\bibnamefont {Patterson}}, \bibinfo {author}
  {\bibfnamefont {G.}~\bibnamefont {Agrawal}}, \bibinfo {author} {\bibfnamefont
  {R.}~\bibnamefont {Bajwa}}, \bibinfo {author} {\bibfnamefont
  {S.}~\bibnamefont {Bates}}, \bibinfo {author} {\bibfnamefont
  {S.}~\bibnamefont {Bhatia}}, \bibinfo {author} {\bibfnamefont
  {N.}~\bibnamefont {Boden}}, \bibinfo {author} {\bibfnamefont
  {A.}~\bibnamefont {Borchers}}, \bibinfo {author} {\bibfnamefont
  {R.}~\bibnamefont {Boyle}}, \bibinfo {author} {\bibfnamefont {P.-l.}\
  \bibnamefont {Cantin}}, \bibinfo {author} {\bibfnamefont {C.}~\bibnamefont
  {Chao}}, \bibinfo {author} {\bibfnamefont {C.}~\bibnamefont {Clark}},
  \bibinfo {author} {\bibfnamefont {J.}~\bibnamefont {Coriell}}, \bibinfo
  {author} {\bibfnamefont {M.}~\bibnamefont {Daley}}, \bibinfo {author}
  {\bibfnamefont {M.}~\bibnamefont {Dau}}, \bibinfo {author} {\bibfnamefont
  {J.}~\bibnamefont {Dean}}, \bibinfo {author} {\bibfnamefont {B.}~\bibnamefont
  {Gelb}}, \bibinfo {author} {\bibfnamefont {T.~V.}\ \bibnamefont
  {Ghaemmaghami}}, \bibinfo {author} {\bibfnamefont {R.}~\bibnamefont
  {Gottipati}}, \bibinfo {author} {\bibfnamefont {W.}~\bibnamefont {Gulland}},
  \bibinfo {author} {\bibfnamefont {R.}~\bibnamefont {Hagmann}}, \bibinfo
  {author} {\bibfnamefont {C.~R.}\ \bibnamefont {Ho}}, \bibinfo {author}
  {\bibfnamefont {D.}~\bibnamefont {Hogberg}}, \bibinfo {author} {\bibfnamefont
  {J.}~\bibnamefont {Hu}}, \bibinfo {author} {\bibfnamefont {R.}~\bibnamefont
  {Hundt}}, \bibinfo {author} {\bibfnamefont {D.}~\bibnamefont {Hurt}},
  \bibinfo {author} {\bibfnamefont {J.}~\bibnamefont {Ibarz}}, \bibinfo
  {author} {\bibfnamefont {A.}~\bibnamefont {Jaffey}}, \bibinfo {author}
  {\bibfnamefont {A.}~\bibnamefont {Jaworski}}, \bibinfo {author}
  {\bibfnamefont {A.}~\bibnamefont {Kaplan}}, \bibinfo {author} {\bibfnamefont
  {H.}~\bibnamefont {Khaitan}}, \bibinfo {author} {\bibfnamefont
  {D.}~\bibnamefont {Killebrew}}, \bibinfo {author} {\bibfnamefont
  {A.}~\bibnamefont {Koch}}, \bibinfo {author} {\bibfnamefont {N.}~\bibnamefont
  {Kumar}}, \bibinfo {author} {\bibfnamefont {S.}~\bibnamefont {Lacy}},
  \bibinfo {author} {\bibfnamefont {J.}~\bibnamefont {Laudon}}, \bibinfo
  {author} {\bibfnamefont {J.}~\bibnamefont {Law}}, \bibinfo {author}
  {\bibfnamefont {D.}~\bibnamefont {Le}}, \bibinfo {author} {\bibfnamefont
  {C.}~\bibnamefont {Leary}}, \bibinfo {author} {\bibfnamefont
  {Z.}~\bibnamefont {Liu}}, \bibinfo {author} {\bibfnamefont {K.}~\bibnamefont
  {Lucke}}, \bibinfo {author} {\bibfnamefont {A.}~\bibnamefont {Lundin}},
  \bibinfo {author} {\bibfnamefont {G.}~\bibnamefont {MacKean}}, \bibinfo
  {author} {\bibfnamefont {A.}~\bibnamefont {Maggiore}}, \bibinfo {author}
  {\bibfnamefont {M.}~\bibnamefont {Mahony}}, \bibinfo {author} {\bibfnamefont
  {K.}~\bibnamefont {Miller}}, \bibinfo {author} {\bibfnamefont
  {R.}~\bibnamefont {Nagarajan}}, \bibinfo {author} {\bibfnamefont
  {R.}~\bibnamefont {Narayanaswami}}, \bibinfo {author} {\bibfnamefont
  {R.}~\bibnamefont {Ni}}, \bibinfo {author} {\bibfnamefont {K.}~\bibnamefont
  {Nix}}, \bibinfo {author} {\bibfnamefont {T.}~\bibnamefont {Norrie}},
  \bibinfo {author} {\bibfnamefont {M.}~\bibnamefont {Omernick}}, \bibinfo
  {author} {\bibfnamefont {N.}~\bibnamefont {Penukonda}}, \bibinfo {author}
  {\bibfnamefont {A.}~\bibnamefont {Phelps}}, \bibinfo {author} {\bibfnamefont
  {J.}~\bibnamefont {Ross}}, \bibinfo {author} {\bibfnamefont {M.}~\bibnamefont
  {Ross}}, \bibinfo {author} {\bibfnamefont {A.}~\bibnamefont {Salek}},
  \bibinfo {author} {\bibfnamefont {E.}~\bibnamefont {Samadiani}}, \bibinfo
  {author} {\bibfnamefont {C.}~\bibnamefont {Severn}}, \bibinfo {author}
  {\bibfnamefont {G.}~\bibnamefont {Sizikov}}, \bibinfo {author} {\bibfnamefont
  {M.}~\bibnamefont {Snelham}}, \bibinfo {author} {\bibfnamefont
  {J.}~\bibnamefont {Souter}}, \bibinfo {author} {\bibfnamefont
  {D.}~\bibnamefont {Steinberg}}, \bibinfo {author} {\bibfnamefont
  {A.}~\bibnamefont {Swing}}, \bibinfo {author} {\bibfnamefont
  {M.}~\bibnamefont {Tan}}, \bibinfo {author} {\bibfnamefont {G.}~\bibnamefont
  {Thorson}}, \bibinfo {author} {\bibfnamefont {B.}~\bibnamefont {Tian}},
  \bibinfo {author} {\bibfnamefont {H.}~\bibnamefont {Toma}}, \bibinfo {author}
  {\bibfnamefont {E.}~\bibnamefont {Tuttle}}, \bibinfo {author} {\bibfnamefont
  {V.}~\bibnamefont {Vasudevan}}, \bibinfo {author} {\bibfnamefont
  {R.}~\bibnamefont {Walter}}, \bibinfo {author} {\bibfnamefont
  {W.}~\bibnamefont {Wang}}, \bibinfo {author} {\bibfnamefont {E.}~\bibnamefont
  {Wilcox}}, \ and\ \bibinfo {author} {\bibfnamefont {D.~H.}\ \bibnamefont
  {Yoon}},\ }in\ \href {\doibase 10.1145/3079856.3080246} {\emph {\bibinfo
  {booktitle} {Proceedings of the 44th Annual International Symposium on
  Computer Architecture}}},\ \bibinfo {series and number} {ISCA '17}\ (\bibinfo
   {publisher} {ACM},\ \bibinfo {address} {New York, NY, USA},\ \bibinfo {year}
  {2017})\ pp.\ \bibinfo {pages} {1--12}\BibitemShut {NoStop}%
\bibitem [{\citenamefont {Blankenbecler}\ \emph {et~al.}(1981)\citenamefont
  {Blankenbecler}, \citenamefont {Scalapino},\ and\ \citenamefont
  {Sugar}}]{Blankenbecler1981}%
  \BibitemOpen
  \bibfield  {author} {\bibinfo {author} {\bibfnamefont {R.}~\bibnamefont
  {Blankenbecler}}, \bibinfo {author} {\bibfnamefont {D.~J.}\ \bibnamefont
  {Scalapino}}, \ and\ \bibinfo {author} {\bibfnamefont {R.~L.}\ \bibnamefont
  {Sugar}},\ }\href {\doibase 10.1103/PhysRevD.24.2278} {\bibfield  {journal}
  {\bibinfo  {journal} {Phys. Rev. D}\ }\textbf {\bibinfo {volume} {24}},\
  \bibinfo {pages} {2278} (\bibinfo {year} {1981})}\BibitemShut {NoStop}%
\bibitem [{\citenamefont {Hirsch}(1985)}]{Hirsch1985}%
  \BibitemOpen
  \bibfield  {author} {\bibinfo {author} {\bibfnamefont {J.~E.}\ \bibnamefont
  {Hirsch}},\ }\href {\doibase 10.1103/PhysRevB.31.4403} {\bibfield  {journal}
  {\bibinfo  {journal} {Phys. Rev. B}\ }\textbf {\bibinfo {volume} {31}},\
  \bibinfo {pages} {4403} (\bibinfo {year} {1985})}\BibitemShut {NoStop}%
\bibitem [{\citenamefont {White}\ \emph {et~al.}(1989)\citenamefont {White},
  \citenamefont {Scalapino}, \citenamefont {Sugar}, \citenamefont {Loh},
  \citenamefont {Gubernatis},\ and\ \citenamefont {Scalettar}}]{White1989}%
  \BibitemOpen
  \bibfield  {author} {\bibinfo {author} {\bibfnamefont {S.~R.}\ \bibnamefont
  {White}}, \bibinfo {author} {\bibfnamefont {D.~J.}\ \bibnamefont
  {Scalapino}}, \bibinfo {author} {\bibfnamefont {R.~L.}\ \bibnamefont
  {Sugar}}, \bibinfo {author} {\bibfnamefont {E.~Y.}\ \bibnamefont {Loh}},
  \bibinfo {author} {\bibfnamefont {J.~E.}\ \bibnamefont {Gubernatis}}, \ and\
  \bibinfo {author} {\bibfnamefont {R.~T.}\ \bibnamefont {Scalettar}},\ }\href
  {\doibase 10.1103/PhysRevB.40.506} {\bibfield  {journal} {\bibinfo  {journal}
  {Phys. Rev. B}\ }\textbf {\bibinfo {volume} {40}},\ \bibinfo {pages} {506}
  (\bibinfo {year} {1989})}\BibitemShut {NoStop}%
\bibitem [{SM()}]{SM}%
  \BibitemOpen
  \href@noop {} {}\bibinfo {note} {See Supplemental Material for (i) proof of
  the relation between the acceptance rate and the training MSE; (ii) basics of
  neural networks; (iii) details of the neural network structure and the
  training hyperparameters; (iv) measured fermion imaginary-time Green's
  function.}\BibitemShut {Stop}%
\bibitem [{\citenamefont {Anderson}(1961)}]{Anderson1961}%
  \BibitemOpen
  \bibfield  {author} {\bibinfo {author} {\bibfnamefont {P.~W.}\ \bibnamefont
  {Anderson}},\ }\href {\doibase 10.1103/PhysRev.124.41} {\bibfield  {journal}
  {\bibinfo  {journal} {Phys. Rev.}\ }\textbf {\bibinfo {volume} {124}},\
  \bibinfo {pages} {41} (\bibinfo {year} {1961})}\BibitemShut {NoStop}%
\bibitem [{\citenamefont {Haldane}(1978)}]{Haldane1978}%
  \BibitemOpen
  \bibfield  {author} {\bibinfo {author} {\bibfnamefont {F.~D.~M.}\
  \bibnamefont {Haldane}},\ }\href {\doibase 10.1103/PhysRevLett.40.416}
  {\bibfield  {journal} {\bibinfo  {journal} {Phys. Rev. Lett.}\ }\textbf
  {\bibinfo {volume} {40}},\ \bibinfo {pages} {416} (\bibinfo {year}
  {1978})}\BibitemShut {NoStop}%
\bibitem [{\citenamefont {Hirsch}\ and\ \citenamefont
  {Fye}(1986)}]{Hirsch1986}%
  \BibitemOpen
  \bibfield  {author} {\bibinfo {author} {\bibfnamefont {J.~E.}\ \bibnamefont
  {Hirsch}}\ and\ \bibinfo {author} {\bibfnamefont {R.~M.}\ \bibnamefont
  {Fye}},\ }\href {\doibase 10.1103/PhysRevLett.56.2521} {\bibfield  {journal}
  {\bibinfo  {journal} {Phys. Rev. Lett.}\ }\textbf {\bibinfo {volume} {56}},\
  \bibinfo {pages} {2521} (\bibinfo {year} {1986})}\BibitemShut {NoStop}%
\bibitem [{\citenamefont {Fye}\ and\ \citenamefont {Hirsch}(1988)}]{Fye1988}%
  \BibitemOpen
  \bibfield  {author} {\bibinfo {author} {\bibfnamefont {R.~M.}\ \bibnamefont
  {Fye}}\ and\ \bibinfo {author} {\bibfnamefont {J.~E.}\ \bibnamefont
  {Hirsch}},\ }\href {\doibase 10.1103/PhysRevB.38.433} {\bibfield  {journal}
  {\bibinfo  {journal} {Phys. Rev. B}\ }\textbf {\bibinfo {volume} {38}},\
  \bibinfo {pages} {433} (\bibinfo {year} {1988})}\BibitemShut {NoStop}%
\bibitem [{\citenamefont {Carleo}\ and\ \citenamefont
  {Troyer}(2017)}]{Carleo2017}%
  \BibitemOpen
  \bibfield  {author} {\bibinfo {author} {\bibfnamefont {G.}~\bibnamefont
  {Carleo}}\ and\ \bibinfo {author} {\bibfnamefont {M.}~\bibnamefont
  {Troyer}},\ }\href {\doibase 10.1126/science.aag2302} {\bibfield  {journal}
  {\bibinfo  {journal} {Science}\ }\textbf {\bibinfo {volume} {355}},\ \bibinfo
  {pages} {602} (\bibinfo {year} {2017})}\BibitemShut {NoStop}%
\bibitem [{\citenamefont {Chen}\ \emph {et~al.}(2018)\citenamefont {Chen},
  \citenamefont {Cheng}, \citenamefont {Xie}, \citenamefont {Wang},\ and\
  \citenamefont {Xiang}}]{Chen2017}%
  \BibitemOpen
  \bibfield  {author} {\bibinfo {author} {\bibfnamefont {J.}~\bibnamefont
  {Chen}}, \bibinfo {author} {\bibfnamefont {S.}~\bibnamefont {Cheng}},
  \bibinfo {author} {\bibfnamefont {H.}~\bibnamefont {Xie}}, \bibinfo {author}
  {\bibfnamefont {L.}~\bibnamefont {Wang}}, \ and\ \bibinfo {author}
  {\bibfnamefont {T.}~\bibnamefont {Xiang}},\ }\href {\doibase
  10.1103/PhysRevB.97.085104} {\bibfield  {journal} {\bibinfo  {journal} {Phys.
  Rev. B}\ }\textbf {\bibinfo {volume} {97}},\ \bibinfo {pages} {085104}
  (\bibinfo {year} {2018})}\BibitemShut {NoStop}%
\bibitem [{\citenamefont {Deng}\ \emph {et~al.}(2017)\citenamefont {Deng},
  \citenamefont {Li},\ and\ \citenamefont {Das~Sarma}}]{Deng2017}%
  \BibitemOpen
  \bibfield  {author} {\bibinfo {author} {\bibfnamefont {D.-L.}\ \bibnamefont
  {Deng}}, \bibinfo {author} {\bibfnamefont {X.}~\bibnamefont {Li}}, \ and\
  \bibinfo {author} {\bibfnamefont {S.}~\bibnamefont {Das~Sarma}},\ }\href
  {\doibase 10.1103/PhysRevX.7.021021} {\bibfield  {journal} {\bibinfo
  {journal} {Phys. Rev. X}\ }\textbf {\bibinfo {volume} {7}},\ \bibinfo {pages}
  {021021} (\bibinfo {year} {2017})}\BibitemShut {NoStop}%
\bibitem [{\citenamefont {Gao}\ and\ \citenamefont {Duan}(2017)}]{Gao2017}%
  \BibitemOpen
  \bibfield  {author} {\bibinfo {author} {\bibfnamefont {X.}~\bibnamefont
  {Gao}}\ and\ \bibinfo {author} {\bibfnamefont {L.-M.}\ \bibnamefont {Duan}},\
  }\href {\doibase 10.1038/s41467-017-00705-2} {\bibfield  {journal} {\bibinfo
  {journal} {Nat. Commun.}\ }\textbf {\bibinfo {volume} {8}},\ \bibinfo {pages}
  {662} (\bibinfo {year} {2017})}\BibitemShut {NoStop}%
\bibitem [{\citenamefont {Huang}\ and\ \citenamefont {Moore}()}]{Huang2017b}%
  \BibitemOpen
  \bibfield  {author} {\bibinfo {author} {\bibfnamefont {Y.}~\bibnamefont
  {Huang}}\ and\ \bibinfo {author} {\bibfnamefont {J.~E.}\ \bibnamefont
  {Moore}},\ }\href {http://arxiv.org/abs/1701.06246} {\ }\Eprint
  {http://arxiv.org/abs/1701.06246} {arXiv:1701.06246} \BibitemShut {NoStop}%
\bibitem [{\citenamefont {Torlai}\ \emph {et~al.}(2018)\citenamefont {Torlai},
  \citenamefont {Mazzola}, \citenamefont {Carrasquilla}, \citenamefont
  {Troyer}, \citenamefont {Melko},\ and\ \citenamefont {Carleo}}]{Torlai2017}%
  \BibitemOpen
  \bibfield  {author} {\bibinfo {author} {\bibfnamefont {G.}~\bibnamefont
  {Torlai}}, \bibinfo {author} {\bibfnamefont {G.}~\bibnamefont {Mazzola}},
  \bibinfo {author} {\bibfnamefont {J.}~\bibnamefont {Carrasquilla}}, \bibinfo
  {author} {\bibfnamefont {M.}~\bibnamefont {Troyer}}, \bibinfo {author}
  {\bibfnamefont {R.}~\bibnamefont {Melko}}, \ and\ \bibinfo {author}
  {\bibfnamefont {G.}~\bibnamefont {Carleo}},\ }\href {\doibase
  10.1038/s41567-018-0048-5} {\bibfield  {journal} {\bibinfo  {journal} {Nat.
  Phys.}\ }\textbf {\bibinfo {volume} {14}},\ \bibinfo {pages} {447} (\bibinfo
  {year} {2018})}\BibitemShut {NoStop}%
\bibitem [{\citenamefont {Saito}\ and\ \citenamefont
  {Kato}(2018)}]{Hiroki2017}%
  \BibitemOpen
  \bibfield  {author} {\bibinfo {author} {\bibfnamefont {H.}~\bibnamefont
  {Saito}}\ and\ \bibinfo {author} {\bibfnamefont {M.}~\bibnamefont {Kato}},\
  }\href {\doibase 10.7566/JPSJ.87.014001} {\bibfield  {journal} {\bibinfo
  {journal} {J. Phys. Soc. Jpn.}\ }\textbf {\bibinfo {volume} {87}},\ \bibinfo
  {pages} {014001} (\bibinfo {year} {2018})}\BibitemShut {NoStop}%
\bibitem [{\citenamefont {Nomura}\ \emph {et~al.}(2017)\citenamefont {Nomura},
  \citenamefont {Darmawan}, \citenamefont {Yamaji},\ and\ \citenamefont
  {Imada}}]{Nomura2017}%
  \BibitemOpen
  \bibfield  {author} {\bibinfo {author} {\bibfnamefont {Y.}~\bibnamefont
  {Nomura}}, \bibinfo {author} {\bibfnamefont {A.~S.}\ \bibnamefont
  {Darmawan}}, \bibinfo {author} {\bibfnamefont {Y.}~\bibnamefont {Yamaji}}, \
  and\ \bibinfo {author} {\bibfnamefont {M.}~\bibnamefont {Imada}},\ }\href
  {\doibase 10.1103/PhysRevB.96.205152} {\bibfield  {journal} {\bibinfo
  {journal} {Phys. Rev. B}\ }\textbf {\bibinfo {volume} {96}},\ \bibinfo
  {pages} {205152} (\bibinfo {year} {2017})}\BibitemShut {NoStop}%
\bibitem [{\citenamefont {Clark}(2018)}]{Clark2017}%
  \BibitemOpen
  \bibfield  {author} {\bibinfo {author} {\bibfnamefont {S.~R.}\ \bibnamefont
  {Clark}},\ }\href {\doibase 10.1088/1751-8121/aaaaf2} {\bibfield  {journal}
  {\bibinfo  {journal} {J. Phys. A Math. Theor.}\ }\textbf {\bibinfo {volume}
  {51}},\ \bibinfo {pages} {135301} (\bibinfo {year} {2018})}\BibitemShut
  {NoStop}%
\bibitem [{\citenamefont {Glasser}\ \emph {et~al.}(2018)\citenamefont
  {Glasser}, \citenamefont {Pancotti}, \citenamefont {August}, \citenamefont
  {Rodriguez},\ and\ \citenamefont {Cirac}}]{Glasser2017}%
  \BibitemOpen
  \bibfield  {author} {\bibinfo {author} {\bibfnamefont {I.}~\bibnamefont
  {Glasser}}, \bibinfo {author} {\bibfnamefont {N.}~\bibnamefont {Pancotti}},
  \bibinfo {author} {\bibfnamefont {M.}~\bibnamefont {August}}, \bibinfo
  {author} {\bibfnamefont {I.~D.}\ \bibnamefont {Rodriguez}}, \ and\ \bibinfo
  {author} {\bibfnamefont {J.~I.}\ \bibnamefont {Cirac}},\ }\href {\doibase
  10.1103/PhysRevX.8.011006} {\bibfield  {journal} {\bibinfo  {journal} {Phys.
  Rev. X}\ }\textbf {\bibinfo {volume} {8}},\ \bibinfo {pages} {011006}
  (\bibinfo {year} {2018})}\BibitemShut {NoStop}%
\bibitem [{\citenamefont {Gull}\ \emph {et~al.}(2007)\citenamefont {Gull},
  \citenamefont {Werner}, \citenamefont {Millis},\ and\ \citenamefont
  {Troyer}}]{Gull2007}%
  \BibitemOpen
  \bibfield  {author} {\bibinfo {author} {\bibfnamefont {E.}~\bibnamefont
  {Gull}}, \bibinfo {author} {\bibfnamefont {P.}~\bibnamefont {Werner}},
  \bibinfo {author} {\bibfnamefont {A.}~\bibnamefont {Millis}}, \ and\ \bibinfo
  {author} {\bibfnamefont {M.}~\bibnamefont {Troyer}},\ }\href {\doibase
  10.1103/PhysRevB.76.235123} {\bibfield  {journal} {\bibinfo  {journal} {Phys.
  Rev. B}\ }\textbf {\bibinfo {volume} {76}},\ \bibinfo {pages} {235123}
  (\bibinfo {year} {2007})}\BibitemShut {NoStop}%
\bibitem [{\citenamefont {Gull}\ \emph {et~al.}(2011)\citenamefont {Gull},
  \citenamefont {Millis}, \citenamefont {Lichtenstein}, \citenamefont
  {Rubtsov}, \citenamefont {Troyer},\ and\ \citenamefont {Werner}}]{Gull2011}%
  \BibitemOpen
  \bibfield  {author} {\bibinfo {author} {\bibfnamefont {E.}~\bibnamefont
  {Gull}}, \bibinfo {author} {\bibfnamefont {A.~J.}\ \bibnamefont {Millis}},
  \bibinfo {author} {\bibfnamefont {A.~I.}\ \bibnamefont {Lichtenstein}},
  \bibinfo {author} {\bibfnamefont {A.~N.}\ \bibnamefont {Rubtsov}}, \bibinfo
  {author} {\bibfnamefont {M.}~\bibnamefont {Troyer}}, \ and\ \bibinfo {author}
  {\bibfnamefont {P.}~\bibnamefont {Werner}},\ }\href {\doibase
  10.1103/RevModPhys.83.349} {\bibfield  {journal} {\bibinfo  {journal} {Rev.
  Mod. Phys.}\ }\textbf {\bibinfo {volume} {83}},\ \bibinfo {pages} {349}
  (\bibinfo {year} {2011})}\BibitemShut {NoStop}%
\bibitem [{\citenamefont {Arsenault}\ \emph {et~al.}(2014)\citenamefont
  {Arsenault}, \citenamefont {Lopez-Bezanilla}, \citenamefont {von
  Lilienfeld},\ and\ \citenamefont {Millis}}]{Arsenault2014}%
  \BibitemOpen
  \bibfield  {author} {\bibinfo {author} {\bibfnamefont {L.-F.}\ \bibnamefont
  {Arsenault}}, \bibinfo {author} {\bibfnamefont {A.}~\bibnamefont
  {Lopez-Bezanilla}}, \bibinfo {author} {\bibfnamefont {O.~A.}\ \bibnamefont
  {von Lilienfeld}}, \ and\ \bibinfo {author} {\bibfnamefont {A.~J.}\
  \bibnamefont {Millis}},\ }\href {\doibase 10.1103/PhysRevB.90.155136}
  {\bibfield  {journal} {\bibinfo  {journal} {Phys. Rev. B}\ }\textbf {\bibinfo
  {volume} {90}},\ \bibinfo {pages} {155136} (\bibinfo {year}
  {2014})}\BibitemShut {NoStop}%
\bibitem [{\citenamefont {Arsenault}\ \emph {et~al.}()\citenamefont
  {Arsenault}, \citenamefont {von Lilienfeld},\ and\ \citenamefont
  {Millis}}]{Arsenault2015}%
  \BibitemOpen
  \bibfield  {author} {\bibinfo {author} {\bibfnamefont {L.-F.}\ \bibnamefont
  {Arsenault}}, \bibinfo {author} {\bibfnamefont {O.~A.}\ \bibnamefont {von
  Lilienfeld}}, \ and\ \bibinfo {author} {\bibfnamefont {A.~J.}\ \bibnamefont
  {Millis}},\ }\href {http://arxiv.org/abs/1506.08858} {\ }\Eprint
  {http://arxiv.org/abs/1506.08858} {arXiv:1506.08858} \BibitemShut {NoStop}%
\bibitem [{\citenamefont {Georges}\ \emph {et~al.}(1996)\citenamefont
  {Georges}, \citenamefont {Kotliar}, \citenamefont {Krauth},\ and\
  \citenamefont {Rozenberg}}]{Georges1996}%
  \BibitemOpen
  \bibfield  {author} {\bibinfo {author} {\bibfnamefont {A.}~\bibnamefont
  {Georges}}, \bibinfo {author} {\bibfnamefont {G.}~\bibnamefont {Kotliar}},
  \bibinfo {author} {\bibfnamefont {W.}~\bibnamefont {Krauth}}, \ and\ \bibinfo
  {author} {\bibfnamefont {M.~J.}\ \bibnamefont {Rozenberg}},\ }\href {\doibase
  10.1103/RevModPhys.68.13} {\bibfield  {journal} {\bibinfo  {journal} {Rev.
  Mod. Phys.}\ }\textbf {\bibinfo {volume} {68}},\ \bibinfo {pages} {13}
  (\bibinfo {year} {1996})}\BibitemShut {NoStop}%
\bibitem [{\citenamefont {Kondor}\ \emph {et~al.}()\citenamefont {Kondor},
  \citenamefont {Son}, \citenamefont {Pan}, \citenamefont {Anderson},\ and\
  \citenamefont {Trivedi}}]{Kondor2018}%
  \BibitemOpen
  \bibfield  {author} {\bibinfo {author} {\bibfnamefont {R.}~\bibnamefont
  {Kondor}}, \bibinfo {author} {\bibfnamefont {H.~T.}\ \bibnamefont {Son}},
  \bibinfo {author} {\bibfnamefont {H.}~\bibnamefont {Pan}}, \bibinfo {author}
  {\bibfnamefont {B.}~\bibnamefont {Anderson}}, \ and\ \bibinfo {author}
  {\bibfnamefont {S.}~\bibnamefont {Trivedi}},\ }\href
  {http://arxiv.org/abs/1801.02144} {\ }\Eprint
  {http://arxiv.org/abs/1801.02144} {arXiv:1801.02144} \BibitemShut {NoStop}%
\end{thebibliography}%

\widetext
\clearpage


\section*{Supplementary material}

# Supplemental Material for "Self-learning Monte Carlo with Deep Neural Networks"

Huitao Shen,[1, *] Junwei Liu,[1, 2, †] and Liang Fu[1]

[1]*Department of physics, Massachusetts Institute of Technology, Cambridge, Massachusetts 02139, USA*
[2]*Department of Physics, Hong Kong University of Science and Technology, Clear Water Bay, Hong Kong, China*

## I. RELATION BETWEEN MSE AND THE ACCEPTANCE RATE

Define the weight difference between the original model and the effective model as $\Delta(\mathcal{S}) \equiv \ln W[S] - \ln W_{\text{eff}}[\mathcal{S}]$, where we have suppressed the $\alpha$ index. Suppose the MSE is minimized as $\min_\alpha \mathbb{E}_{\mathcal{S} \sim W[\mathcal{S}]/Z} \Delta(\mathcal{S})^2 \equiv l^2$. The expectation of the logarithmic acceptance rate is estimated as [1]

$$\mathbb{E}_{\mathcal{S}_1 \sim W[\mathcal{S}_1]/Z, \mathcal{S}_2 \sim W[\mathcal{S}_2]/Z} \left[\ln p(\mathcal{S}_1 \to \mathcal{S}_2)\right]^2 \tag{1}$$

$$= \mathbb{E}_{\mathcal{S}_1 \sim W[\mathcal{S}_1]/Z, \mathcal{S}_2 \sim W[\mathcal{S}_2]/Z} \left[\min(0, \Delta(\mathcal{S}_2) - \Delta(\mathcal{S}_1))\right]^2 \tag{2}$$

$$= \mathbb{E}_{\mathcal{S}_1 \sim W[\mathcal{S}_1]/Z, \mathcal{S}_2 \sim W[\mathcal{S}_2]/Z} \frac{1}{2} \left[\Delta(\mathcal{S}_2) - \Delta(\mathcal{S}_1)\right]^2 \tag{3}$$

$$= \mathbb{E}_{\mathcal{S} \sim W[\mathcal{S}]/Z} \Delta(\mathcal{S})^2 \tag{4}$$

$$= l^2. \tag{5}$$

From Eq. (2) to Eq. (3), we note the fact that half of the configurations have $\Delta(\mathcal{S}_2) - \Delta(\mathcal{S}_1) < 0$. From Eq. (3) to Eq. (4) we have assumed $\mathbb{E}_{\mathcal{S} \sim W[\mathcal{S}]/Z} \Delta(\mathcal{S}) = 0$, meaning our effective model is unbiased. Hence we obtain $\langle (\ln p)^2 \rangle = l^2 = \text{MSE}$.

If the effective model is biased with $\mathbb{E}_{\mathcal{S} \sim W[\mathcal{S}]/Z} \Delta(\mathcal{S}) = \mu$, the formula becomes $\langle (\ln p)^2 \rangle = \text{MSE} - \mu^2$. This is a manifestation of the bias–variance tradeoff in statistics.

Compared with the coefficient of determination $R^2$ defined in Ref. [2], this quantity is directly related to the acceptance rate, which further determines the efficiency of SLMC. It is more meaningful to use this quantity to characterize how well the effective model resembles the original model.

## II. FULLY-CONNECTED NEURAL NETWORKS

*Mathematical Description* A fully-connected neural network is composed of several fully-connected layers. The effect of $i$-th fully-connected layer can be summarized as $\mathbf{a}_{i+1} = f_i(W_i \mathbf{a}_i + \mathbf{b}_i)$, where $\mathbf{a}_i$ is the input/output vector of the $i$-th/$(i-1)$-th layer. $W_i$, $\mathbf{b}_i$, $f_i$ are the weight matrix, bias vector, and the nonlinear activation function of such layer. The layer is said to have $N_i$ neurons when $W_i$ is of size $N_i \times N_{i-1}$.

*Training Data* The training set is small compared with that of the whole configuration space. In order for the neural network to perform well, the limited training data should to be as typical as possible. Therefore they are generated by importance sampling from the Monte Carlo simulation of the original model. Also, these data are sufficiently independent from each other.

The training data $\{\mathbf{s}, \ln W[\mathbf{s}]\}$ are all fed directly into the neural network without normalization. The input auxiliary spins are $\pm 1$, which are already pretty well-conditioned. We do not normalize $\ln W[\mathbf{s}]$ (typically of order 10) because the third layer is a linear layer which should automatically scale the data.

The size of the training set is 30000, which is always larger than the trainable parameters to prevent over-fitting. The size of the test set is 10000.

*Training Hyperparameters* The neural networks are trained with TensorFlow [3] using back-propagation algorithm. All the network training in this paper were done in a laptop, and every training task was completed within 10 minutes.

We use mini-batch training with batch size 128. All the training is completed using the Adam optimizer [4] with a learning rate 0.001. Xavier initialization [5] for the weights, and constant initialization to 0.01 for the biases are adopted. Since our networks are relatively small, the loss would converge to a satisfactory level in several hundreds of epochs for a training starting from scratch. Typical loss during the training can be found in Fig. 1. In order to

---

[*] huitao@mit.edu
[†] liuj@ust.hk

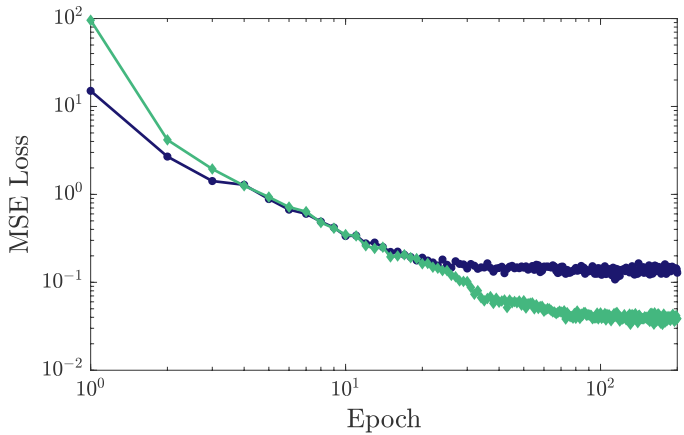

FIG. 1. Typical MSE loss for a fully-connected network of $\mu = -1$ in Fig. 1 in the main text (blue, circle) and for a convolutional network of $\mu = -1$ in Fig. 4 in the main text (green, diamond) during a training instance.

TABLE I. Performances of the fully-connected neural networks of different sizes. Here $\beta = 20, U = 3.0, \mu = -1, V = 1.0$ and $L = 120$.

| $(N_1, N_2)$ | Trainables[*] | MSE($\times 10^{-1}$) | Acceptance Rate |
|---|---|---|---|
| (140,70) | 26881 | 1.31 | 73.06 |
| (120,60) | 21841 | 1.22 | 72.92 |
| (100,50) | 17201 | 1.33 | 73.34 |
| (80,40) | 12961 | 1.54 | 68.94 |
| (60,30) | 9121 | 1.87 | 66.40 |
| (40,20) | 5681 | 2.55 | 61.90 |

[*]"Trainables" is the number of trainable parameters in the neural network. Same abbreviation for Table. II.

obtain good visualization, proper choice of regularization is necessary. We always use both $L_1$ and $L_2$ regularization with strength 0.005 to encourage sparsity.

*Iterative Training* When computing Fig. 1 in the main text, we used the technique of iterative training. Instead of training a new effective model from scratch for every chemical potential $\mu$, we only do so once (for example, for the $\mu = 0$ model). For other chemical potentials, we train effective models using the already trained model as the starting point. The losses would then converge typically in less than 5 epoches. This technique is particularly useful when a lot of similar models need to be simulated systematically.

*Performance Benchmark* In benchmarking the performance of the neural network, the MSE is evaluated on a test set of independent configurations. The acceptance rate is averaged over 5000 consecutive proposals of global moves. Each global move involves $10L$ local updates on the effective model so that the proposed configuration is uncorrelated with the original one.

*Effect of Network Size* The performance of the neural networks with respect to the network size is summarized in Table. I. For simplicity, we always fix $N_2 = N_1/2$. As long as $L \sim N_1$, the MSEs could be minimized to a reasonable value. More complex neural networks could not enhance the performance significantly, possibly due to the increasing redundancy in large networks.

*Effect of Activation Function* We have tested the three activation functions: rectified linear unit (ReLU), sigmoid, tanh.

$$\text{ReLU: } f(x) = \max\{x, 0\}, \tag{6}$$

$$\text{sigmoid: } f(x) = \frac{1}{1 + e^{-x}}. \tag{7}$$

$$\text{tanh: } f(x) = \tanh x = \frac{1 - e^{-2x}}{1 + e^{-2x}}. \tag{8}$$

The performance of the effective models are not very sensitive to the choice of activation functions. Nevertheless, we empirically found ReLU neural networks are much easier to train especially in the empty orbital regime, i.e., small $\mu$.

TABLE II. Structures of the convolutional networks involved in this paper. For convolutional layers, $(n,d,t)$ is the (kernel number, kernel size, stride); For fully-connected layers, $N$ is the number of neurons.

| $\beta$ | 1st | 2nd | 3rd | 4th | Trainables |
|---|---|---|---|---|---|
| 10 | (6,11,5) | (2,4,2) | 6 | - | 195 |
| 20 | (2/6,11,5) | (2,4,2) | 7 | - | 211/291 |
| 30 | (6,11,5) | (2,4,2) | 7 | - | 375 |
| 40 | (6,11,5) | (2,6,2) | (2,4,2) | 7 | 319 |
| 50 | (6,11,5) | (2,6,2) | (2,6,2) | 7 | 355 |

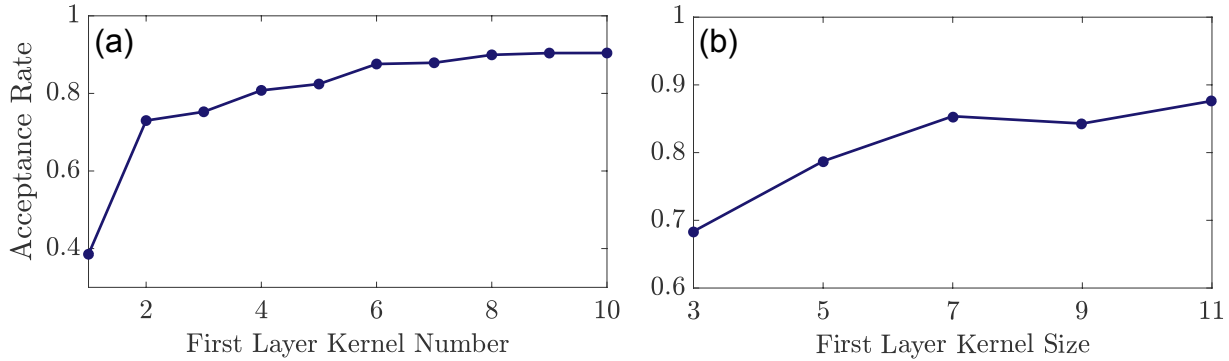

FIG. 2. The effect of (a) the kernel number and (b) the kernel size of the first convolutional layer in the network on the performance of the effective model. The second layer is fixed to have 2 kernels of size 4 and stride 2. Here $\beta = 20, U = 3.0, \mu = -1, V = 1.0$ and $L = 120$. For (a), the size of the kernel is fixed to be 11 and the stride is 5. For (b), the size of the final fully-connected layer is adjusted correspondingly.

## III. CONVOLUTIONAL NEURAL NETWORKS

*Mathematical Description* The convolutional neural network used in this paper is composed of several convolutional layers and several fully-connected layers.

The input and the output of each convolutional layer, instead of being vectors, are now generally matrices. The input vector of auxiliary spin can be seen as a $L \times 1$ matrix. The large weight matrix is replaced by several small matrices called "kernels". The effect of the convolutional layer is the sliding inner product (also called cross-correlation) between the kernels $h$ and the input matrix $a$, which is defined as

$$[a * h]_{i,j} \equiv [a * h^j]_i \equiv \sum_{k=1}^{d} \sum_{l=1}^{a_y} a_{t(i-1)+k,l} h^j_{k,l}, \qquad (9)$$

where $h^j$ is the $j$-th kernel. The size of the kernel is $d$, and the stride of the sliding inner product is $t$. $l$ is the number of kernels in the previous convolutional layer. We take the periodic boundary condition (also called padding) for the sliding inner product (not shown in Fig. 3 in the main text for simplicity). Suppose there are $n$ kernels, and the dimension of the input matrix is $a_x \times a_y$, then the dimensions of the kernels should be $d \times a_y$, and the dimension of the output matrix is $a_x/t \times n$. The effect of $i$-th such layer can be denoted compactly as $a_{i+1} = f_i(a_i * h_i + b_i)$. $b_i$ is the bias associated with kernel $h^i$.

After propagating the input through several convolutional layers, the matrix is flattened into a vector so that it can be used as the input of subsequent fully-connected layers. In the literature, this operation is often formally performed by a "flatten layer" in between the convolutional layers and the fully-connected layers.

*Structures of the Networks* See Table. II. The activation functions for all convolutional networks are ReLU.

*Training Hyperparameters* The size of the data set, the initialization, the regularization and the learning rate are all the same as those of the fully-connected network, except that there is no $L_1$ regularization for convolutional networks. Typical loss during the training can be found in Fig. 1. Iterative training is also used in computing Fig. 4(a) in the main text.

We note that for convolutional layers, there exists possibility that the network was stuck at some local minima of the loss function, possibly because the landscape of the loss function is very drastic due to the small number of

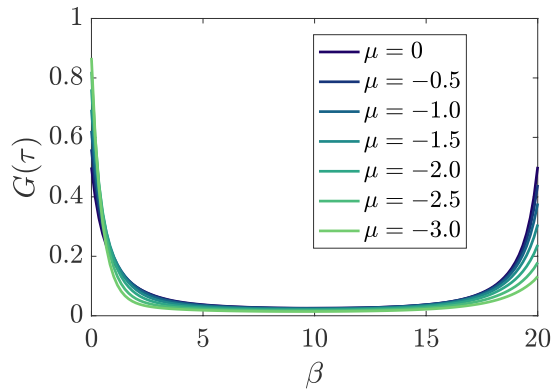

FIG. 3. Measured impurity electron Green's function $G(\tau)$ for different chemical potentials $\mu$. The measurement is averaged over 5000 independent configurations of auxiliary spins. Here $\beta = 20, U = 3.0, V = 1.0$ and $L = 2\beta U$.

trainable parameters. One should always try to train the network for two or three times and pick the best one on the validation set.

*Effect of Kernel Number* The performance of the convolutional networks with respect to the kernel size is shown in Fig. 2(a). Here only the kernel number of the first layer is varied. As can be seen, the performance decreases significantly when the number is smaller than two, which is in accordance with Fig. 2 in the main text that there are mainly two features in the auxiliary spin configuration. The performance would be slightly improved if the kernel number is further increased beyond two.

*Effect of Kernel Size* The performance of the convolutional networks with respect to the kernel size is shown in Fig. 2(b). Here only the kernel size of the first layer is varied. Large kernel size leads to better acceptance rates, probably because of the fewer trainable parameters in these networks.

## IV. MEASURED GREEN'S FUNCTION

The Green's function of impurity electrons is defined as

$$G_{\sigma\sigma'}(\tau - \tau') = \langle T_\tau \hat{d}_\sigma(\tau) \hat{d}^\dagger_{\sigma'}(\tau') \rangle_\beta, \tag{10}$$

where $\langle \cdots \rangle_\beta \equiv \text{Tr}[e^{-\beta \hat{H}} \cdots]/\text{Tr}[e^{-\beta \hat{H}}]$ is the thermal ensemble average, $T_\tau$ is the time-ordering operator, and $\hat{d}_\sigma$ is the impurity electron annihilation operator (see Eq. (4)-(6) in the main text). In the definition, we have already exploited the translational symmetry in the imaginary time. Since spin is conserved in the asymmetric Anderson model, we have $G_{\uparrow\downarrow}(\tau) = G_{\downarrow\uparrow}(\tau) = 0$ and $G(\tau) \equiv G_{\uparrow\uparrow}(\tau) = G_{\downarrow\downarrow}(\tau)$. In Fig. 3 we show measured $G(\tau)$ using SLMC.

[1] We would like to thank an anonymous referee for pointing out a mistake in our previous derivation.
[2] J. Liu, H. Shen, Y. Qi, Z. Y. Meng, and L. Fu, Phys. Rev. B **95**, 241104 (2017).
[3] M. Abadi, A. Agarwal, P. Barham, E. Brevdo, Z. Chen, C. Citro, G. S. Corrado, A. Davis, J. Dean, M. Devin, S. Ghemawat, I. Goodfellow, A. Harp, G. Irving, M. Isard, Y. Jia, R. Jozefowicz, L. Kaiser, M. Kudlur, J. Levenberg, D. Mané, R. Monga, S. Moore, D. Murray, C. Olah, M. Schuster, J. Shlens, B. Steiner, I. Sutskever, K. Talwar, P. Tucker, V. Vanhoucke, V. Vasudevan, F. Viégas, O. Vinyals, P. Warden, M. Wattenberg, M. Wicke, Y. Yu, and X. Zheng, "TensorFlow: Large-scale machine learning on heterogeneous systems," (2015), software available from tensorflow.org.
[4] D. P. Kingma and J. Ba, in *International Conference on Learning Representations (ICLR2015)*, arXiv:1412.6980.
[5] X. Glorot and Y. Bengio, in *Proceedings of the Thirteenth International Conference on Artificial Intelligence and Statistics*, Proceedings of Machine Learning Research, Vol. 9, edited by Y. W. Teh and M. Titterington (PMLR, Chia Laguna Resort, Sardinia, Italy, 2010) pp. 249–256.



\end{document}